\documentclass[twocolumn,showpacs,pre,amsmath,amssymb,floatfix]{revtex4}

\usepackage{amsmath,amssymb}
\usepackage{bm}
\usepackage{epsfig}

\newcommand{\bd}{\bm}

\begin{document}

\title{Nonlocal effective average action approach to crystalline phantom membranes}
\author{N. Hasselmann$^{1,2}$}
\author{F. L. Braghin$^{2,3}$} 
\date{December 1, 2010}

\affiliation{$^1$Max-Planck-Institute for Solid State Research, Heisenbergstr.~1, D-70569 Stuttgart, Germany\\
$^2$International Institute of Physics, Univ. Fed. do Rio
Grande do Norte, 59072-970, Natal, RN, Brazil  
\\
$^3$Instituto de F\'\i sica, Univ. Fed. de Goi\'as, P.B.131, Campus II, 
74001-970, Goi\^ania, GO, Brazil
}

\begin{abstract}
We investigate the properties of crystalline phantom membranes,
 at the crumpling
transition and in the flat phase, using a nonperturbative renormalization
group approach. We avoid a derivative expansion of the effective average
action and instead analyse the full momentum dependence of the elastic
coupling functions. This leads to a more accurate determination
of the critical exponents and further yields the full momentum dependence
of the correlation functions of the in-plane and out-of-plane fluctuation.
The flow equations are solved numerically 
for $D=2$ dimensional membranes embedded in a $d=3$ dimensional
space.
Within our approach we find a crumpling transition of second order 
which is
characterized by an anomalous exponent $\eta_c\approx 0.63(8)$ and
the thermal exponent $\nu\approx 0.69$. Near the crumpling transition
the order parameter of the flat phase vanishes with a critical exponent
$\beta\approx 0.22$.
The flat phase anomalous dimension
is $\eta_f\approx 0.85$ and the Poisson's ratio inside
the flat phase is found to be $\sigma_f\approx -1/3$. At the crumpling
transition we find a much larger negative value of the Poisson's ratio
$\sigma_c \approx -0.71(5)$. 
We discuss further in detail the different
regimes of the momentum dependent fluctuations, both in the flat phase
and in the vicinity of the crumpling transition, and extract the 
crossover momentum scales which separate them.
\end{abstract}

\maketitle

\section{Introduction}
Crystalline membranes have a fixed connectivity of their
constituent particles and therefore a finite in-plane shear modulus which
distinguishes them from fluid membranes \cite{membranebook}. 
Physical crystalline membranes are
two dimensional membranes embedded in three dimensional space and
the embedding allows for out-of-plane fluctuations into the third dimension
which are absent in a truly two-dimensional crystal. The coupling
of the in-plane modes to the out-of-plane ones is responsible for
a strong renormalization of the infrared (IR) character of both
modes. Both acquire large anomalous dimensions
which render the fluctuation of the local normals of
the membrane finite and thus stabilize a flat phase in which the
normals are ordered \cite{Nelson87}. 
The physics of such flat crystalline membranes 
has recently received renewed attention
because of the discovery of graphene \cite{Castro09}, 
the thinnest possible crystalline
membrane \cite{Meyer07}.
At higher temperatures, 
the flat phase eventually becomes unstable and looses its orientational
order of the normals at the crumpling transition \cite{Paczuski88}.
Several approaches have been used to investigate the crumpling 
transition of 
phantom membranes. In contrast to real membranes, phantom
membranes have no self-avoiding
contact interaction which makes them more accessible to analytical
approaches.
Paczuski {\em et al.} studied a generalized model 
of a $D$-dimensional elastic manifold embedded in a $d$-dimensional
space within a leading order 
$\epsilon=4-D$ expansion and found for $D=2$ ($\epsilon=2$) 
a first order transition for all $d<219$. The accuracy of this
approach, which is well controlled only for small $\epsilon$,
at $\epsilon=2$ is however in doubt. The results from
the $\epsilon$ expansion 
disagree for $D=2$ with the results
of a large $d$ expansion \cite{Paczuski89,Aronovitz89} and its extension, the 
selfconsistent screening approximation (SCSA) \cite{Doussal92,Gazit09,Zakharchenko10}
where a continuous
transition was found for $D=2$ and $d=3$. Further support
for a continuous transition came from an elastic model
with a constraint arising from infinitely large 
coupling constants  for the stretching modes \cite{David88}, from a 
Monte Carlo (MC)
renormalization group analysis \cite{Espriu96}, and from early
MC simulations, see references in \cite{Bowick01rev}. More
recent MC simulations found however evidence for the
coexistence of two separate phases
at the critical temperature which would be consistent with a first
order transition \cite{Kownacki02,Koibuchi04}, see also
\cite{Koibuchi08,Nishiyama10}. 
A recent nonperturbative renormalization group (NPRG) analysis, 
based on a derivative expansion
of the effective average action, could reproduce both the leading order
results of the $\epsilon=4-D$ expansion as well as the leading
order of a large $d$ expansion, within a unified framework 
\cite{Kownacki09}. The technique has recently also been applied
to anisotropic membranes \cite{Essafi10}.
For
$D=2$ and $d=3$ evidence of a second order phase transition was
found,
but a weak first order transition could not be ruled out
because of a weak dependence of the flow on the employed cutoff scheme.

Crystalline membranes are characterized by an anomalous elasticity
in which none of the local elastic constants remain finite. 
The elasticity becomes fully nonlocal
in the IR limit and the elastic coupling constants become coupling
functions with a momentum dependence which for asymptotically small
momenta is characterized by anomalous exponents. 
Here we present a NPRG analysis where the full momentum dependence
of the coupling functions is kept, which allows for a more accurate
analysis of the crumpling transition which we find to be
continous. Our approach further allows
to calculate the
thermal fluctuations of the membrane beyond the asymptotic
regime. 
We recover the results of 
Ref.~\cite{Kownacki09} if the momentum dependence of the
coupling functions is neglected.
We previously applied our approach
to compute the thermal fluctuations of free standing graphene \cite{Braghin10}
and found excellent agreement for all momenta
with MC simulations \cite{Fasolino07,Los09}. 
This analysis
allowed to identify the characteristic scale of ripples in free standing
graphene \cite{Meyer07} with the Ginzburg scale of the nonlinear elastic theory
of crystalline membranes. 
The SCSA is also capable of accessing fluctuations
at finite momenta and was applied to investigate the
thermal fluctuations of graphene \cite{Zakharchenko10}, but proved
to be, at finite
momenta, less accurate than the NPRG approach.
Here, we extend our approach and
investigate in detail the behavior of a crystalline membrane
inside the flat phase and at the crumpling transition
critical point. 

All approximations of our approach
are included in our nonlocal ansatz of the effective average
action and not at the level of the flow equations for irreducible
vertices. Once the effective average action, which
respects the full symmetry of the symmetric phase, is specified,
the flow equations of the irreducible vertices are uniquely
determined and can be solved exactly.
In all previous schemes to calculate the full momentum dependence 
of self-energies of classical models
(see Refs.~\cite{phi4} for approaches for the $\Phi^4$ theory), some approximations were done in addition
to the truncation of the effective average action which in principle can
lead to a violation of Ward identities.
Our approach is quite general and could also be applied to 
other
physical models.
Similar flow equations
were derived for a model of interacting bosons \cite{Sinner10}. However, additional
approximations had to be employed to solve the flow equations in which case
the approach becomes equivalent to that presented in \cite{Sinner09}.
The approach presented here assumes
an ordered state but allows to approach and analyse
the critical point from within the
low temperature phase.

The structure of this manuscript is the following. In Sec.~\ref{sec:model}
we introduce and define the usual Landau-Ginzburg model of crystalline phantom
membranes, which are membranes with no self-avoiding
contact interaction. In Sec.~\ref{sec:NPRG} we present the derivation
of our NPRG flow equations for the nonlocal elastic coupling functions.
Expressions for the two longest diagrams of the flow equations can be found
in the appendix.
In Sec.~\ref{sec:results} we present results from a completely
self-consistent numerical solution of the flow equations. We analyse the
different scaling regimes in the momentum dependence of the out-of-plane
and in-plane fluctuations and determine the critical exponents
of the membrane both in the flat phase and at the crumpling transition
fixed point. We further discuss the Poisson's ratio, which is different
at the crumpling transition and inside the flat phase.
Finally, in Sec.~\ref{sec:conclusion} we present a summary and conclusion
of this work.

\section{Landau-Ginzburg model}
\label{sec:model}
Our starting point is the usual Landau-Ginzburg model for crystalline
phantom membranes (which have no contact interaction) 
\cite{Paczuski88}, with ${\cal H}={\cal H}^{\rm b}+{\cal H}^{\rm st}$. 
The bending part of the membranes is described by (silent
indices are summed over)
\begin{subequations}
\begin{equation}
{\cal H}^{\rm b}=\frac{\tilde\kappa}{2}\int d^D{x} \left(\partial_a\partial_a {\bd R}\right)^2
\label{eq:bend}
\end{equation}
and the stretching is described by
\begin{eqnarray}
{\cal H}^{\rm st}&=&\int d^D x \big[\frac{{\tilde r}_0}{2}(\partial_a {\bd R})^2+
\frac{\tilde{\mu}}{4}
(\partial_a {\bd R}\cdot \partial_b {\bd R})^2 \nonumber \\
&&+\frac{\tilde{\lambda}}{8}
(\partial_a {\bd R}\cdot \partial_a {\bd R})^2 \big] \, ,
\label{eq:stretch}
\end{eqnarray}
\end{subequations}
where ${\bd R}$ is a $D+1$ dimensional vector parameterizing the $D$
dimensional membrane which is embedded in a $d=D+1$ dimensional space. 
For notational
convenience we restrict our analysis to $d=D+1$, it is however straightforward
to extend our analysis to arbitrary $d$.
In terms of the derivatives ${\bd m}_a=\partial_a {\bd R}$ the 
stretching part becomes
\begin{eqnarray}
{\cal H}^{\rm st} &=& \int d^D x \left[ \frac{\tilde{r}_0}{2}  
\bm{m}_a \cdot \bm{m}_a +
\frac{\tilde{\mu}}{4}  
(\bm{m}_a \cdot \bm{m}_b)^2 \right.
\nonumber
\\
&& +  \left. \frac{{\tilde{\lambda}}}{8} \left(
\bm{m}_a \cdot \bm{m}_a \right)^2 \right] \, ,
\label{Htanvariables}
\end{eqnarray}
and one can already anticipate a transition near $\tilde{r}_0\simeq 0$ from
a flat configuration, which exists for negative $\tilde{r}_0$ and where the 
fields ${\bd m}_a$ acquire a finite expectation value 
${\bd m}_{a,0}=\big< \bd{m}_a \big>\neq 0$, to a crumpled phase with
$\big< \bd{m}_a \big>=0$. The stretching part Eq.~(\ref{Htanvariables})
is appropriate for an isotropic membrane whose consituents particles
have a fixed connectivity (such membranes are also refered to as
tethered or polymerized membranes). A crystalline membrane has in general
a lower symmetry and additional terms quartic in $\bd{m}_a$ are allowed,
except for a two-dimensional hexagonal lattice \cite{Aronovitz89}. 
We here concentrate
on models described by  Eq.~(\ref{Htanvariables}), appropriate for isotropic
manifolds or membranes with hexagonal lattices and $D=2$.

For the flat phase, we introduce $J\neq 0$ as the magnitude
of the order parameter which is defined as 
$\big< \bd{R} (\bd{x}) \big> = J x_a \bd{e}_a$ with $a=1\dots D$
and $\bd{e}_a$ are orthonormal vectors which span the plane of the membrane. 
Defining $U_{ab}=(\bd{m}_a \cdot \bd{m}_b-\bd{m}_{a,0} \cdot \bd{m}_{b,0})/2$
and $\bd{m}_{a,0}=J \bd{e}_a$,
one can write \cite{Aronovitz89}
\begin{align}
  {\cal H}^{\rm st}&=\int d^Dx \Big[\tilde{\mu}\, 
  U_{ab}^2 +\frac{\tilde{\lambda}}{2} \,  
  U_{aa}^2
  \Big] 
  \nonumber \\
  & \qquad +  
  \int d^Dx\Big({\tilde r}_0+J^2[\tilde{\mu}+D\tilde{\lambda}/2]\Big)
  U_{aa} \, .
  \label{eq:flatstretch}
\end{align}
The last term has to vanish to guarantee thermodynamic stability and
we thus have the mean field result 
\begin{equation}
  \label{eq:Jmf}
  J_{\Lambda_0}^2=-\tilde{r}_0/[\tilde{\mu}_{\Lambda_0}+D \tilde{\lambda}_{\Lambda_0}/2] \, ,
\end{equation}
where the subscript $\Lambda_0$ indicates that this is a relation among the bare 
parameters which are assumed to be defined at some scale $\Lambda_0$.

\section{Derivation of the nonperturbative renormalization group flow equations}
\label{sec:NPRG}
The NPRG approach is based on the exact flow equation of the 
effective average action $\Gamma_\Lambda$ \cite{Wetterich93}
\begin{equation}
  \label{eq:wett}
  \frac{\partial \Gamma_\Lambda}{\partial \Lambda}=
  \frac{1}{2} {\rm Tr} \left[ \left( \frac{\partial^2\Gamma_\Lambda}{\partial \phi \partial \phi^\prime}
      +R_\Lambda \right)^{-1} \frac{\partial R_\Lambda}{\partial \Lambda} \right]
\end{equation}
where the fields $\phi,\phi^\prime$ here represent the components
of the vector field ${\bd R}$
and the trace stands for a momentum integral
and a sum over internal indices.
The function  $R_\Lambda$ is a regulator that removes IR divergences
 arising from modes with $k < \Lambda$ and will be specified below.

In principle, since the NPRG is not based on an expansion in
a small parameter, the NPRG is not a-priori controlled. It has
however been applied with much success to critical phenomena
\cite{Berges02} and it usually reproduces leading order
results of perturbative RG techniques such as 
$\epsilon$ expansions around the upper (or lower) critical
dimension. For the case of crystalline membranes, the simplest
possible NPRG ansatz \cite{Kownacki09} 
already reproduces the leading order results
of both an $\epsilon=4-D$ expansion as well as
a $1/d$ expansion. The approach presented here 
allows to calculate the momentum dependence of the membrane fluctuations
which were found to be in perfect agreement with large scale MC
simulations of unsuspended graphene \cite{Braghin10}.
The NPRG technique thus seems to be quite reliable when applied to
crystalline membranes.

\subsection{Nonlocal elasticity}
\label{sec:nonlocal}
Our approximation of the effective average action consists of a 
nonlocal bending part $\Gamma^{\rm b}$, which is quadratic 
in the fields, and a
nonlinear stretching part $\Gamma^{\rm st}$, which also includes 
quartic terms,
\begin{eqnarray}
  \Gamma_\Lambda &=& \Gamma_\Lambda^{\rm b} +  \Gamma_\Lambda^{\rm st} \nonumber 
  \\
  &=&
  \frac{1}{2} \int d^Dx \ d^D x^\prime {\tilde \kappa}_\Lambda({\bd x}-{\bd x}^\prime)
\partial_a^2 {\bd R}({\bd x}) \partial_b^2 {\bd R}({\bd x}^\prime)
\nonumber
\\ && +
\int d^Dx \ d^Dx^\prime \Big[ {\tilde \mu}_\Lambda({\bd x}-{\bd x}^\prime)
U_{ab}({\bd x})U_{ab}({\bd x}^\prime) \nonumber
\\ && \qquad +\frac{1}{2}{\tilde \lambda}_\Lambda({\bd x}-{\bd x}^\prime) 
U_{aa}({\bd x})U_{bb}({\bd x}^\prime) \Big] .
\label{eq:nonloc-ansatz}
\end{eqnarray}
This ansatz is simply a nonlocal generalization of the bare model 
defined via
Eqs.~(\ref{eq:bend}) and (\ref{eq:flatstretch}).
Nonlocal correlations are known to be dominant 
in the IR limit of crystalline
membranes where the Fourier-transformed coupling functions 
scale anomalously. 
Both in the flat phase and at the critical
point of the crumpling transition,
one has a pronounced hardening
of the out-of plane fluctuations and a softening of the in-plane fluctuations.
This is expressed by a divergent
form of the bending coupling function and a 
vanishing of the bulk and shear modulus.
In the flat phase one has 
$\tilde{\kappa}(q)\sim q^{-\eta_f}$ whereas the elastic coupling functions vanish as
$\tilde{\mu}(q)\sim\tilde{\lambda}(q)\sim q^{\eta^u}$. Here, $\eta_f$
and $\eta^u$ are 
the anomalous exponents of the out-of plane fluctuations 
and the in-plane fluctuations, respectively. 
Note that these anomalous exponents are in fact not independent. 
Invariance of the original model under rotations 
implies a relation (Ward identity) among the
anomalous dimensions
\cite{Aronovitz89}, 
\begin{equation}
  \eta^u=4-D-2 \eta_f \, .
  \label{eq:ward}
\end{equation}
Since all fluctuations become anomalous in the IR limit, 
neither $\tilde{\kappa}(q)$, $\tilde{\mu}(k)$, nor
$\tilde{\lambda}(k)$ are analytic functions in the limit $\Lambda\to 0$
and all fluctuations become nonlocal, a situation
familiar from the behavior at a critical point of a continuous
phase transition. In a leading order derivative expansion only the
momentum independent parts of the coupling functions 
are kept, which
is sufficient to extract their asymptotic momentum
dependence by approximating $f_{\Lambda=0}(k)\approx f_{\Lambda=k}(0)$
where $f$ is any of the three coupling functions.
We choose a more accurate approach, which also captures the corrections
to the asymptotic behavior, and simply keep
the full momentum dependence of the coupling functions for all $\Lambda$.

It is convenient to write the flow equation
(\ref{eq:wett}) of the functional $\Gamma_\Lambda$ as a flow equation
of vertex functions, which can be obtained from a field expansion
of both sides of (\ref{eq:wett}).
Since we are interested in the symmetry broken phase, we work with fields
corresponding to the in-plane fluctuations ${\bd u}$ and out-of-plane
fluctuations $h$ rather than the original $\bd{R}$ fields. 
These are introduced via the fluctuations of
$\bd{m}_a$ around the ordered state,
$\Delta{\bd m}_a
=\partial_a\bd{R}-\bd{m}_{a,0}$ with $\Delta{\bd m}_a
=(\partial_a {\bd u},\partial_a h)$
and ${\bd m}_{a,0}=J_\Lambda {\bd e}_a$.
We therefore write the expansion
\begin{multline}
\Gamma_\Lambda [{\bd u},h]=
\sum_{n,m=0}^\infty\frac{1}{n! m!} \iint \limits_{\substack{{\bd p}_1\dots {\bd p}_m 
\\ {\bd q}_1\dots {\bd q}_n}} (2\pi)^D \delta \left(
\sum_{i=1}^n{\bd p}_i+
\sum_{i=1}^m{\bd q}_i \right)  
\\ 
\times \sum_{a_1 \dots a_m=1}^D
\Gamma^{(n+m)}_{\Lambda,{\underbrace{h\dots h}_{\mbox{\tiny n times}}}
a_1\dots a_m}
({\bd q}_1, \dots ,{\bd q}_n ;{\bd p}_1, \dots ,{\bd p}_m ) \\ \times
h_{\bd{q}_1} \dots h_{\bd{q}_n}
u_{\bd{p}_1}^{a_1} \dots u_{\bd{p}_m}^{a_m} \, ,
\label{eq:gammaexp}
\end{multline}
where the momentum integrals are defined as
\begin{equation} 
  \int_{{\bm q}} = \int \frac{d^D q}{(2 \pi)^D} \, .
\end{equation}

The irreducible vertices entering Eq.~(\ref{eq:gammaexp}) can
be related to correlation functions. The simplest example
is the Dyson equation which relates the one particle Green's function
to the irreducible self-energies.
The Dyson equation for the Green's function $G_{hh}$ of the $h$ field, 
with
\begin{equation}
  \big< h_{\bd q} h_{-{\bd q}^\prime}\big>=
  V \delta_{{\bd q},{\bd q^\prime}} G_{hh}(q) \, ,
\end{equation} 
where $V$ is the $D$-dimensional volume, is
\begin{equation}
G_{hh}^{-1}(q)=G_{0,\Lambda}^{-1}(q)+\Sigma_{hh}(q) \, .
\end{equation}
The self-energy 
is of the form (here and below we suppress in our notation the
explicit
$\Lambda$ dependence of the coupling parameters $\tilde{\kappa}_q$
$\tilde{\mu}_q$, and $\tilde{\lambda}_q$)
\begin{equation}
  \Sigma_{hh}(q)=\Gamma^{(2)}_{hh}({\bd q},-{\bd q})=
  ({\tilde \kappa}_q-{\tilde \kappa}_{\Lambda_0})q^4 , 
\end{equation}
and the cutoff dependent non-interacting Green's function is
\begin{equation}
  G_{0,\Lambda}^{-1}(q)={\tilde \kappa}_{\Lambda_0} q^4+R_\Lambda (q),
\end{equation}
where ${\tilde \kappa}_{\Lambda_0}$ denotes the bare and momentum independent value of the
initial coupling constant ${\tilde \kappa}$ defined at the ultraviolet (UV) 
cutoff 
$\Lambda_0$. Here, we introduced the regulator function $R_\Lambda(q)$,
which regulates the IR limit of the propagator in such a way that
the IR divergence at $q\to 0$ is removed at finite $\Lambda$.
Since we will later solve the NPRG flow equations numerically, we
choose an analytic cutoff \cite{Kownacki09},
\begin{equation}
  R_\Lambda(q)= \tilde{\kappa}_{\Lambda}^{(0)} 
\frac{q^4}{\exp[(q/\Lambda)^4]-1} \, ,
\end{equation}
where 
\begin{equation}
  \tilde{\kappa}_\Lambda^{(0)}=\tilde{\kappa}_{q=0}
\end{equation}
is the $q=0$ component of the Fourier
transform of $\tilde{\kappa}_\Lambda({\bd x})$.
For finite $\Lambda$, we then have  
$\lim_{q\to 0}G_{0,\Lambda}^{-1}=\tilde{\kappa}_{\Lambda}^{(0)}\Lambda^4$ 
and $G_{0,\Lambda}$
is non-divergent for $q\to 0$.

The Green's functions of the in-plane modes, defined via
\begin{equation}
  \big<u^a_{\bd k} u^b_{-{\bd k}^\prime}\big>=
  V \delta_{{\bd k},{\bd k}^\prime} G_{ab}({\bd k}),
\end{equation}
can be written in terms of transverse ($\perp$) and longitudinal ($\parallel$) 
components using
the projectors
\begin{subequations}
  \begin{align}
    P^\perp_{ab}({\bm k}) &= \delta_{ab}-k^a k^b/k^2 \, , \\
    P^\parallel_{ab}({\bm k}) &= k^a k^b/k^2 \, .
  \end{align}
\end{subequations}
This yields
\begin{equation}
G_{ab}({\bd k}) = G_\perp(k) P^\perp_{ab}({\bm k})
+ G_\parallel(k) P^\parallel_{ab}({\bm k})
\end{equation}
\noindent with
$G_{\alpha}^{-1}= G_{0,\Lambda}^{-1}+\Sigma_\alpha$ for $\alpha=\perp,\parallel$ and
where the self-energies are defined as projections of
\begin{equation}
\Sigma_{ab}({\bd k}) = \Gamma_{ab}^{(2)}({\bd k},-\bd{k}) \, .
\end{equation}
With the two-point irreducible vertex 
\begin{align}
\Gamma_{ab}^{(2)} (\bd{k}_1, -\bd{k}_2) 
&=J_\Lambda^2 \Big\{
  \tilde\mu_{k_1} \big(\delta_{ab}
{\bd k}_1 \cdot {\bd k}_2 +k_2^a k_1^b\big) +
\tilde{\lambda}_{k_1} k_1^a k_2^b \Big\}
\nonumber \\ & \quad
+(\tilde{\kappa}_{k_1}-\tilde{\kappa}_{\Lambda_0}) k_1^2 k_2^2 \, ,
\end{align}
one finds the in-plane self-energies
\begin{subequations}
  \begin{eqnarray}
    \label{eq:sigperp}
    \Sigma_\perp(k)&=&J_\Lambda^2 {\tilde \mu}_k k^2+({\tilde \kappa}_k-{\tilde \kappa}_{\Lambda_0})k^4 , \\
    \Sigma_\parallel(k)&=&J_\Lambda^2 (2{\tilde \mu}_k+{\tilde \lambda}_k) k^2+
    ({\tilde \kappa}_k-{\tilde \kappa}_{\Lambda_0})k^4 .
    \label{eq:sigpar}
  \end{eqnarray}
\end{subequations}

The effective average action Eq.~(\ref{eq:nonloc-ansatz}) contains further three- and four-point
vertices of the form
(we use $k$ for momenta of $u$-fields and $p$ 
for momenta of $h$-fields)
\begin{subequations}
  \begin{align}
    \label{eq:g3}
    \Gamma^{(3)}_{a_1a_2a_3}(\bd{k}_1,\bd{k}_2,\bd{k}_3)
    &= \frac{-i J_\Lambda}{2} \sum_{(\alpha \beta \gamma)={\cal P}_{123}} \big\{ 
    2 \tilde{\mu}_{k_\alpha}    (\bd{k}_{\alpha}\cdot \bd{k}_\beta)
    k_\gamma^{a_\alpha} \nonumber \\
    & \quad +
    \tilde{\lambda}_{k_\alpha} (\bd{k}_\beta \cdot
    \bd{k}_\gamma) k_{\alpha}^{a_\alpha} 
    \big\}
    \delta_{a_\beta, a_\gamma}
    \\
    \Gamma^{(3)}_{hha}(\bd{p}_1,\bd{p}_2;\bd{k})&=-iJ_\Lambda\big\{
    \tilde{\mu}_k \big[(\bd{p}_1\cdot \bd{k}) p_2^a +(\bd{p}_2\cdot \bd{k} ) p_1^a\big]
    \nonumber \\ & \quad
    + \tilde{\lambda}_k (\bd{p}_1\cdot \bd{p}_2 ) k^a \big\} \, ,
  \end{align}
  \begin{multline}
    \Gamma^{(4)}_{a_1 \dots a_4}(\bd{k}_1 \dots \bd{k}_4)   
    =
    \frac{1}{8}\sum \limits_{\substack{(\alpha \beta \gamma \delta)  
        \\ ={\cal P}_{1234} }}
    \big\{
    2\tilde{\mu}_{k_{\alpha\beta}} 
    (\bd{k}_\alpha\cdot \bd{k}_\gamma)( \bd{k}_\beta\cdot \bd{k}_\delta)
    \\  \, \, \, \qquad
    +\tilde{\lambda}_{k_{\alpha\beta}} (\bd{k}_\alpha\cdot \bd{k}_\beta)
    ( \bd{k}_\gamma\cdot \bd{k}_\delta)\big\} 
    \delta_{a_\alpha, a_\beta}\delta_{a_\gamma, a_\delta}  \, ,
  \end{multline}
  \begin{multline}
    \Gamma^{(4)}_{hha_1a_2}(\bd{p}_1,\bd{p}_2;\bd{k}_1,\bd{k}_2)=
    \big\{ \tilde{\mu}_{p_{12}} \big[(\bd{p}_1 \cdot \bd{k}_1)(\bd{p}_2 \cdot \bd{k}_2) 
    \\
    +
    (\bd{p}_1 \cdot \bd{k}_2)(\bd{p}_2 \cdot \bd{k}_1)\big] +\tilde{\lambda}_{p_{12}} (\bd{p}_1 \cdot \bd{p}_2)
    (\bd{k}_1 \cdot \bd{k}_2)
    \big\} \delta_{a_1,a_2} \, ,
  \end{multline}
  \begin{multline}
    \Gamma^{(4)}_{hhhh}(\bd{p}_1 \dots \bd{p}_4) \\ 
    =\frac{1}{8}\sum \limits_{\substack{(\alpha \beta \gamma \delta)  
        ={\cal P}_{1234} }}
    \big\{ 2\tilde{\mu}_{p_{\alpha\beta}}+\tilde{\lambda}_{p_{\alpha\gamma}} \big\}
    (\bd{p}_\alpha\cdot \bd{p}_\gamma )(\bd{p}_\beta\cdot \bd{p}_\delta) \, ,
    \label{eq:g4}
  \end{multline}
\end{subequations}
where $k_{\alpha \beta}=|{\bd k}_\alpha+{\bd k}_\beta|$ and 
$\sum_{(\alpha \beta \gamma)={\cal P}_{123}}$ denotes a summation
over all permutations of $(123)$.
 The subscript $h$ 
refers to $h$-fields while subscripts $a$ and $a_i$  refer to 
${u}$-fields.  Since we neglect all irreducible correlations not explicitely
included in Eq.~(\ref{eq:nonloc-ansatz}), higher order vertices,
with more than four legs, do not appear in our approximation.  
Finally, we shall also need the single scale propagators
which are defined via 
\begin{subequations}
  \begin{align}
    {\dot G}_{ab}&=-\Big( G_{\perp}^2 P_{ab}^\perp+G_\parallel^2 P_{ab}^\parallel \Big)
    \partial_\Lambda
    R_{\Lambda} \, , \\ 
    {\dot G}_{hh} &= -G_{hh}^2\partial_\Lambda
    R_{\Lambda} \, .
  \end{align}
\end{subequations}

\subsection{Flow equations for momentum dependent  vertices}
\label{sec:floweqs}
We will now derive the flow equations for the order parameter
$J_\Lambda$ and the coupling functions $\tilde{\kappa}_q$,
$\tilde{\mu}_q$ and $\tilde{\lambda}_q$. The flow of
$J_\Lambda^2$ can be extracted
from the flow equation of the one-point vertex of the ${u}$-fields
whereas the other three coupling functions can be extracted
from the flow equations of the self-energies. All higher order
vertices presented in Sec.~\ref{sec:nonlocal} are completely
determined by $J_\Lambda^2$, $\tilde{\kappa}_q$,
$\tilde{\mu}_q$ and $\tilde{\lambda}_q$, and the flow
equations for the one- and two-point vertices are thus
closed.

The only approximation of our approach is the approximation
for the effective average action as expressed through 
Eq.~(\ref{eq:nonloc-ansatz})
and in the derivation below no further approximation is required. 
The resulting flow equations 
are uniquely determined by Eqs.~(\ref{eq:wett}) and (\ref{eq:nonloc-ansatz})
and obey the full symmetry of Eq.~(\ref{eq:nonloc-ansatz}).

The NPRG flow of the one-point function $\Gamma_{a}^{(1)}$
is   
\begin{align} \label{RGflows}
\partial_{\Lambda} \Gamma_{a}^{(1)} ({\bm k}) &= -
 i (\partial_{\Lambda}J_\Lambda) 
(2\pi)^D  
\delta^{(D)}({\bm k})  
\frac{\partial}{\partial {k'}^b} \Gamma^{(2)}_{ab} ({\bm k},-\bm{k}')
\Big|_{\bd{k}'=\bd{k}}
\nonumber
\\
&
+ \frac{1}{2} \int_{{\bm q}} \big\{\dot{G}_{hh}(q)
\Gamma^{(3)}_{hha} ({\bm q}, - {\bm q}; {\bm k}) 
\nonumber
\\
&  \qquad +
\dot{G}_{bc}({\bm q})\Gamma^{(3)}_{abc}
({\bm k},  {\bm q},- {\bm q}) \big\} (2 \pi)^D \delta^{(D)}({\bm k}) \, ,
\end{align}
where we used the $\bd{k}$-space representation of the order parameter 
field 
\begin{equation}
  \big< R^a\big>(\bd{k})=-i J_\Lambda(2\pi)^D \delta^{(D)}(\bd{k})
  \frac{\partial}{\partial {k^a}} \, .
\end{equation}
\begin{figure}[ht]
\includegraphics[width=7.cm]{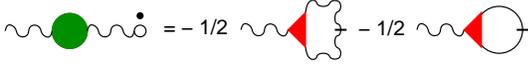}
\caption{(Color online) Diagrammatic representation of the flow equation 
which determines the flow of the order
parameter and which arises from the vanishing flow of
the one-point vertex \cite{Schuetz06}.
Wavy lines correspond to $G_{ab}$ propagators and solid lines to
$G_{hh}$. Lines with a dash correspond to single-scale
propagators. The small open circle with a dot denotes 
a derivative of the order parameter 
with respect to $\Lambda$. } \label{fig:flow1pnt}
\end{figure}

The condition $\partial_\Lambda \Gamma_{a}^{(1)}(\bd{k})=0$ then yields 
an equation which is shown diagrammatically in Fig.~\ref{fig:flow1pnt}
and from which the flow of
the order parameter can be determined. One finds
\begin{multline}
  \partial_{\Lambda} J_\Lambda^2 
  =
  \frac{K_D}{[2 \tilde{\mu}_0 + D {\tilde{\lambda}}_0]D } 
  \int d  q  \; q^{D+1}
  \Big\{
  (D-1) \dot{G}_{\perp} (q)  \\  
  \times      [ 2 \tilde{\mu}_q 
  + 
  2 \tilde{\mu}_0 
  + D {\tilde{\lambda}}_0 
  ]
  \\
  + \dot{G}_{\parallel} (q) \left[ 
    2(2\tilde{\mu}_q + \tilde{\mu}_0) +  
    2 {\tilde{\lambda}}_q
    + D {\tilde{\lambda}}_0   \right] \\
  + \dot{G}_{hh} (q)
  [ 2\tilde{\mu}_0 + D {\tilde{\lambda}}_0 ]
  \Big\} \, ,
\label{eq:J2flow}
\end{multline}
where 
\begin{equation}
  K_D=\frac{ 1}{2^{D-1} \pi^{D/2}\Gamma[D/2]} \, .
\end{equation}

\begin{figure}[ht]
\includegraphics[width=7.cm]{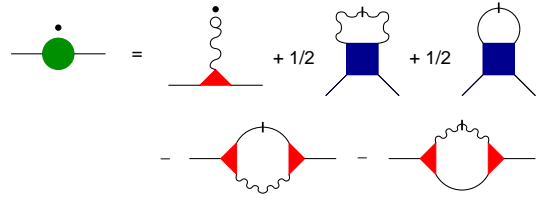}
\caption{(Color online) Diagrams which enter the flow of the self-energy $\Sigma_{hh}$ of the
out-of-plane fluctuations. 
The solid dot above the self-energy on the l.h.s. denotes a
derivative w.r.t. $\Lambda$, the other symbols are defined in the caption
of Fig.~\ref{fig:flow1pnt}.} 
\label{fig:flowsighh}
\end{figure}

The flow of $\tilde{\kappa}(q)$ follows directly 
from the flow equation for $\Sigma_{hh}$. The flow equation is
shown diagrammatically in Fig.~\ref{fig:flowsighh} and is given
by the expression
\begin{align} \label{eq:flowhh}
  \partial_{\Lambda}\Sigma_{hh}(k)&=
k^4  \partial_{\Lambda} {\tilde \kappa}_{k}  \nonumber \\ 
&= k^2 \frac{\partial_{\Lambda} J_\Lambda^2}{2}
  ( 2 \tilde{\mu}_0  + D {\tilde{\lambda}}_0  ) 
  \nonumber \\ & \quad 
  + [ S_{hh}^{h} (k) + S_{hh}^{u} (k) ]/2
  - S_{hh}^{uh}(k) - S_{hh}^{hu} (k) \, .
\end{align}

The diagrams entering the flow of $\Sigma_{hh}(k)$ are 
\begin{subequations}
  \begin{align} \label{diag-h1}    
    S^{h}_{hh} (k) &=
    \int_{{\bm q}}
    \dot{G}_{hh} (q)
    \Gamma_{hhhh}^{(4)} ({\bm k},-{\bm k},{\bm q},-{\bm q}) \, ,
    \\
    S^{u}_{hh} (k) &=
    \int_{{\bm q}}
    \dot{G}_{ab} ({\bm q})
    \Gamma_{hhab}^{(4)} ({\bm k},-{\bm k};{\bm q},-{\bm q}) \, , 
    \\
    S^{hu}_{hh} (k)   &= 
    \int_{{\bm q}} \dot{G}_{hh} ({\bm q}) 
    G_{ab}({\bm q}') \nonumber \\ & \qquad \times
    \Gamma_{hha}^{(3)} ({\bm k},{\bm q};-{\bm q}') 
    \Gamma_{hhb}^{(3)} (-{\bm k},-{\bm q};{\bm q}') \, ,
    \\
    S^{uh}_{hh} (k) &=  
    \int_{{\bm q}}
    \dot{G}_{ab} ({\bm q})  G_{hh} ({\bm q}') \nonumber \\ & \qquad \times
    \Gamma_{hha}^{(3)} ({\bm k},-{\bm q}';{\bm q}) 
    \Gamma_{hhb}^{(3)} (-{\bm k},{\bm q}';-{\bm q}) \, ,
  \end{align}
\end{subequations}
where the direction of ${\bm k}$ on the right hand side of these
equations can be chosen arbitrarily since the diagrams only depend
on the modulus $k$, and ${\bm q}^\prime = {\bm k}+{\bm q}$.
The subscript $hh$ indicates two external $h$-legs and
the superscript indicates the composition of the
internal loop, i.e. $S_{hh}^h$ indicates a diagram where
the loop consists of one $h$-line. $S_{hh}^{uh}$ differs
from $S_{hh}^{hu}$ in that in $S_{hh}^{uh}$ the single-scale
propagator is a $u$-line whereas in $S_{hh}^{hu}$ it is a 
$h$-line.
With the vertices given in Eqs.~(\ref{eq:g3}-\ref{eq:g4}) these
diagrams become
\begin{subequations}
  \begin{align} 
    \label{eq:Shhh}
    S^{h}_{hh} (k) &=  k^2  \int_{\bm q} q^2 
    \dot{G}_{hh} (q) \nonumber \\ & \quad
    \times \big[ 
    2 \tilde{\mu}_{q^{\prime}}  + {\tilde{\lambda}}_0  
    + 2y^2 ( \tilde{\mu}_0 + \tilde{\mu}_{q^{\prime}}   
    + {\tilde{\lambda}}_{q^{\prime}})
    \big] \, ,
    \\
    S^{u}_{hh}(k) &= k^2
    \int_{\bm q} q^2
    [\dot{G}_{\perp} (q) (D-1) + \dot{G}_{\parallel} (q)]
    ( 2 y^2 \tilde{\mu}_0 + {\tilde{\lambda}}_0) \, ,
    \\
    S_{hh}^{hu} (k) &= k^2 J_\Lambda^2  
    \int_{{\bm q}}
    q^2 \dot{G}_{hh}(q')
    \Big\{
    G_{\perp} (q) \tilde{\mu}_q^2 ( 1 - y^2) (2 ky +q)^2 
    \nonumber
    \\
    & \qquad
    + G_{\parallel} (q) 
    \big[ 2 \tilde{\mu}_q y (q+ky)
    +
    \tilde{\lambda}_q (k+ qy) \big]^2
    \Big\} \, ,
    \\
    S_{hh}^{uh} (k) &= k^2 J_\Lambda^2  
    \int_{{\bm q}}
    q^2 G_{hh}(q')
    \Big\{
    \dot{G}_{\perp} (q) \tilde{\mu}_q^2 ( 1 - y^2) (2 ky +q)^2 
    \nonumber
    \\
    & \qquad
    + \dot{G}_{\parallel} (q) 
    \big[ 2 \tilde{\mu}_q y (q+ky)
    +
    \tilde{\lambda}_q (k+ qy) \big]^2
    \Big\} \, . \label{eq:Shhuh}
  \end{align}
\end{subequations}
Note that each of the diagrams given in Eqs.~(\ref{eq:Shhh}-\ref{eq:Shhuh})
is to leading order quadratic in $k$. However, their sum cancels
to leading order exactly the first term in Eq.~(\ref{eq:flowhh}) so
that the overall leading term of the flow of $\Sigma_{hh}$ is indeed
quartic in $k$.

\begin{figure}[ht]
\includegraphics[width=7.cm]{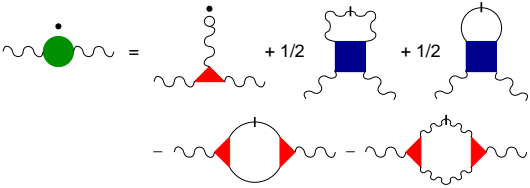}
\caption{(Color online) Diagrams which determine the flow of the self-energy $\Sigma_{ab}$ 
of the in-plane fluctuations. 
The symbols are defined as 
in Figs.~\ref{fig:flow1pnt} and \ref{fig:flowsighh}.} \label{fig:flowsigab}
\end{figure}

Finally, using the flow equation of $\Sigma_{ab}({\bm k})$, one can determine the 
flow of $\tilde{\mu}_k$ and $\tilde{\lambda}_k$. The flow of  $\Sigma_{ab}({\bm k})$ 
has the form (see Fig.~\ref{fig:flowsigab})
\begin{align} \label{eq:flowsigab}
  \partial_{\Lambda} {\Sigma}_{ab}({\bm k}) &=
  \big[ (D\tilde{\lambda}_0+2 \tilde{\mu}_k+2\tilde{\mu}_0)k^2 \delta_{ab}
  \nonumber \\ & \qquad
  +
  2( \tilde{\lambda}_k + \tilde{\mu}_k) k^a k^b \big] \partial_\Lambda J_\Lambda
  +
  (1/2) S_{ab}^{u} ({\bm k}) \nonumber \\
  & \qquad 
  + (1/2) S_{ab}^{h}({\bm k}) 
  - S_{ab}^{hh} ({\bm k})- S_{ab}^{uu}({\bm k}) \, ,
\end{align}
where the first term arises from the diagram with one external
leg coupled to the derivative of the order parameter field and
$S_{ab}^{u} ({\bm k})$ stands for a diagram with external $u$-lines with
flavors
$a,b$ and momentum $k$, and an internal loop which consists of one 
$u$-line.
Similarly, for $S_{ab}^{uu} ({\bm k})$ 
the internal loop consists of two
$u$-lines and for $S_{ab}^{hh} ({\bm k})$ the internal loop consists
of two $h$-lines. 
The analytic expressions for these diagrams are
\begin{subequations}
  \begin{align} \label{diagSk2}
    S_{ab}^{h} ({\bm k}) &=
    \int_{{\bm q}}
    \dot{G}_{hh} (q)
    \Gamma^{(4)}_{hhab}( {\bm k},-{\bm k};{\bm q},-{\bm q}) \, ,
    \\
    S_{ab}^{u} ({\bm k}) &=
    \int_{{\bm q}}
    \dot{G}_{cd} ({\bd q})
    \Gamma_{abcd}^{(4)}({\bm k},-{\bm k},{\bm q},-{\bm q}) \, ,
    \\
    S_{ab}^{hh} ({\bm k}) &=
    \int_{{\bm q}}
    \dot{G}_{hh} (q)G_{hh}(q')  \nonumber
    \\
    & \qquad \times \Gamma_{hha}^{(3)} ({\bm q},-{\bm q}';{\bm k}) 
    \Gamma_{hhb}^{(3)} (-{\bm q},{\bm q}';-{\bm k}) \, ,
    \\
    S_{ab}^{uu}  ({\bm k}) &=
    \int_{{\bm q}}
    \dot{G}_{cd}({\bm q}) {G}_{ef} ({\bm q}')
    \nonumber
    \\
    & \qquad \times \Gamma_{ace}^{(3)}({\bm k},{\bm q},-{\bm q}') 
    \Gamma_{bdf}^{(3)}(-{\bm k},-{\bm q},{\bm q}') \, ,
    \label{eq:Sabuu}
  \end{align}
\end{subequations}
where ${\bm q}^\prime = {\bm k}+{\bm q}$. After projecting on
transversal and longitudinal components, one finds
from Eq.~(\ref{eq:flowsigab}) and the form of the projected
self-energies in Eqs.~(\ref{eq:sigperp}) and (\ref{eq:sigpar})
\begin{subequations}
  \begin{align}
    \label{eq:flowmu}
    k^2\partial_\Lambda \tilde{\mu}_k &=
    k^2(\tilde{\mu}_0+D \tilde{\lambda}_0 /2)\partial_\Lambda J_\Lambda^2 
    \nonumber 
    +(1/2) S_{\perp}^{u} (k)  \\ & \quad
    + (1/2) S_{\perp}^{h}(k)
    - S_{\perp}^{hh} (k)- S_{\perp}^{uu}(k)
    \nonumber \\ & \quad
    -k^4\partial_\Lambda \tilde{\kappa}_k \, , \\
    k^2\partial_\Lambda (2\tilde{\mu}_k+\tilde{\lambda}_k) &=
    k^2(\tilde{\mu}_0+D \tilde{\lambda}_0 /2)\partial_\Lambda J_\Lambda^2 
    \nonumber 
    +(1/2) S_{\parallel}^{u} (k)  \\ & \quad
    + (1/2) S_{\parallel}^{h}(k)
    - S_{\parallel}^{hh} (k)- S_{\parallel}^{uu}(k)
    \nonumber \\ & \quad
    -k^4\partial_\Lambda \tilde{\kappa}_k \, .
    \label{eq:flowmula}
  \end{align} 
\end{subequations}
The projected diagram $S_{\alpha}^{u} (k)$, with $\alpha=\perp$ or $\parallel$,
is defined via $P_{ac}^\alpha ({\bm k})S_{cb}^u({\bm k})=
P_{ab}^\alpha ({\bm k})S_{\alpha}^u(k)$ and the other projected
diagrams are similarly defined. 
With $y={\bm k}\cdot {\bm q}/k q$, the transverse 
projected diagrams have the
form 
\begin{subequations}
  \begin{align}  \label{diagII}
    S_{\perp}^{h}(k)  &= 
    k^2 \int_{{\bm q}} q^2
    \dot{G}_{hh}(q) \big[ 
    2 \tilde{\mu}_0 y^2+ 
    {\tilde{\lambda}}_0 \big] \, ,
    \\
    S_{\perp}^{u}(k) &=
    \frac{k^2}{D-1} \int_{{\bm q}} q^2
    \Big\{
    \big[ \dot{G}_{\perp} (q) (D-1) + \dot{G}_{\parallel} (q) \big]  
    \nonumber \\ & \qquad
    \times 
    (D-1) \big[2  \tilde{\mu}_0 y^2 + 
    {\tilde{\lambda}}_0 \big] \nonumber \\ & \quad
    +\big[ \dot{G}_{\perp} (q) (D - 2 + y^2) 
    +\dot{G}_{\parallel} (q) 
    (1- y^2) \big] 
    \nonumber \\ & \qquad
    \times 2\big[ \tilde{\mu}_{q^\prime} ( 1+y^2 )  
    + {\tilde{\lambda}}_{q^\prime}  y^2
    \big] 
    \Big\} \, ,
    \\
    S_{\perp}^{hh}(k)  &=
    \frac{J_\Lambda^2 k^2}{D-1}
    \int_{{\bm q}} q^2
    \dot{G}_{hh}(q) G_{hh}(q') 
    \nonumber \\ & \qquad \times
    \tilde{\mu}_{k}^2 ( k+ 2 y q)^2   (1-y^2) \, .
  \end{align}
\end{subequations}
The expression of the diagram  $S_{\perp}^{uu}(k)$ 
is rather long and can be found in the appendix.
The longitudinal projections are
\begin{subequations}
  \begin{align}  \label{diagII-t}
    S_{\parallel}^{h}(k) &= 
    S_{\perp}^{h}(k) \, ,
    \\
    S_{\parallel}^{u}(k) &= 
    k^2 \int_{{\bm q}} q^2
    \Big\{
    \big[ \dot{G}_{\perp} (q) (D-1) + \dot{G}_{\parallel} (q) \big]  
    \nonumber \\ & \qquad
    \times 
    (D-1) \big[2  \tilde{\mu}_0 y^2 + 
    {\tilde{\lambda}}_0 \big] \nonumber \\ & \quad
    +\big[ \dot{G}_{\perp} (q) (1- y^2) 
    +\dot{G}_{\parallel} (q) 
    y^2 \big] \nonumber \\ & \qquad
    \times 2\big[ \tilde{\mu}_{q^\prime} ( 1+y^2 )  
    + {\tilde{\lambda}}_{q^\prime}  y^2
    \big] 
    \Big\} \, ,
    \\
    S_{\parallel}^{hh}(k)  &=
    J_\Lambda^2 k^2
    \int_{{\bm q}} q^2
    \dot{G}_{hh}(q) G_{hh}(q') 
    \nonumber \\ & \qquad \times
    \big[\tilde{\lambda}_k (q+k y) + 2 \tilde{\mu}_k  (k+q y)y  
    \big]^2 \, ,
  \end{align}
\end{subequations}
and the expression for $S_\parallel^{uu}(k)$ is given 
in the appendix.

\section{Results}
\label{sec:results}
The set of Eqs.~(\ref{eq:J2flow}), (\ref{eq:flowhh}), (\ref{eq:flowmu}), and (\ref{eq:flowmula}) 
form a set of coupled integrodifferential equations which we solve numerically. 
We are mainly interested in the behavior of the $D=2$ dimensional membrane embedded in three
dimensional space 
near its critical regime. For simplicity we use
as initial conditions momentum independent coupling constants,
\begin{equation}
  \tilde{\kappa}_{\Lambda_0}=1 \, \, \,  \mbox{and} \, \, \, 
  \tilde{\mu}_{\Lambda_0}=\tilde{\lambda}_{\Lambda_0}=\Lambda_0^2 \, .
  \label{eq:initvalues}
\end{equation}
The initial mean field value of the order parameter $J_{\Lambda_0}$ will be tuned to reach the
critical point of the crumpling transition. Since our approach breaks down within the crumpled phase, we can approach
the critical point only from the flat phase. We begin by discussing the properties of the flat
phase. 

\subsection{Flat phase}
The flat phase  is characterized by a finite order parameter $J=\lim_{\Lambda \to 0}J_\Lambda$.
Asymptotically, the out-of-plane fluctuations are governed by the anomalous dimension $\eta_f$
associated with the flat phase fixed point,
\begin{equation}
  G_{hh}(k)\sim q^{-(4-\eta_f)} \, ,
  \label{eq:ghhanom}
\end{equation}
while the in-plane fluctuations are governed by the anomalous dimension $\eta^u$,
\begin{equation}
  G_{\perp}(q) \sim G_{\parallel}(q) \sim q^{-(2+\eta^u)} \, .
  \label{eq:guuanom}
\end{equation}
We can extract the flow of the anomalous dimension from the leading 
$q=0$ term of $\tilde{\kappa}(q)$,
\begin{equation}
 \Lambda \partial_\Lambda \tilde{\kappa}^{(0)}_\Lambda =- \eta_\Lambda 
\tilde{\kappa}^{(0)}_\Lambda \, ,
\end{equation}
where $\tilde{\kappa}_\Lambda^{(0)}=\tilde{\kappa}_{q=0}$.
In accordance with the derivative expansion result
\cite{Kownacki09} and recent MC simulations of graphene \cite{Los09} we find for $D=2$
the anomalous dimension
$\eta_f \approx 0.85$ which yields $\eta^u \approx 0.30$. These values are also
close to the SCSA result $\eta_f\approx 0.821$ \cite{Doussal92} and 
the MC results $\eta\approx 0.750(5)$ \cite{Bowick96} and 
$\eta\approx 0.81(3)$ \cite{Zhang93}. 
A typical flow of the anomalous dimension
far away from the critical point is shown in the upmost curve in Fig.~\ref{fig:eta}. 
There is some scatter of the numerical data for $\eta_\Lambda$ since
$\eta_\Lambda$ has to be extracted from the $\Lambda$ derivative
of the very small $q$ dependence
of $\tilde{\kappa}_q$. The numerical noise in the function $\tilde{\kappa}_q$
itself is however very small.
The anomalous scaling of the propagators, as expressed through 
Eqs.~(\ref{eq:ghhanom})  and ({\ref{eq:guuanom}),
is observable only for momenta smaller than the Ginzburg scale $q_G$. This scale can be obtained 
perturbatively and has for $D=2$ the form
\cite{membranebook,Nelson87} 
\begin{equation}
q_G\approx [3\tilde{K}_0/(2\pi)]^{1/2}/(4 {\tilde \kappa})
\label{eq:ginz}
\end{equation}
where 
\begin{equation}
\tilde{K}_0=4J_{\Lambda_0}^2{\tilde \mu}({\tilde
\mu}+{\tilde \lambda}) /(2 {\tilde \mu}+{\tilde \lambda}) \, 
\end{equation}
is the (bare) Young's modulus.
Since for all our calculations we used Eq.~(\ref{eq:initvalues}) and
similar values of
$J_{\Lambda_0}^2\approx 0.5$, one finds for
all the data we present here to a good approximation a common Ginzburg scale
of the order of $q_G\approx 0.2 \Lambda_0$. For $q\ll q_G$ the Green's functions are strongly
renormalized and the flat phase fixed point scaling regime appears. This behavior is clearly
seen in both $G_{hh}(q)$ and $G_{\perp} (q)$, see the lowest curves 
in Figs.~\ref{fig:ghh} and 
\ref{fig:gperp}, which where obtained for $J_{\Lambda_0}^2=1/2$ or $\delta J=J_{\Lambda_0}-J_c\approx 0.08$,
where $J_c$ is the critical $J_{\Lambda_0}$ value of the crumpling transition for the initial values of the elastic
constants stated in Eq.~(\ref{eq:initvalues}). 
In the small $q$ limit, we find $G_{hh}(q)\propto 1/q^{4-\eta_f}$, as expected
from the analysis of the flow of $\eta_\Lambda$. Similarly, for the in-plane
correlation functions we find $G_\alpha \propto 1/q^{2+\eta^u}$ with 
$\alpha=\perp,\parallel$ and an anomalous exponent $\eta^u=2-2\eta_f$, as
expected from the Ward identity Eq.~(\ref{eq:ward}).

The upper curve in Fig.~\ref{fig:poisson} shows the flow of the Poisson's ratio
\begin{equation}
  \sigma_\Lambda=\frac{\tilde{\lambda}_{q=0}}{2 \tilde{\mu}_{q=0}+ \tilde{\lambda}_{q=0}} \, ,
  \label{eq:poisson}
\end{equation}
which in the flat phase is known to acquire a negative value for 
$\Lambda\to 0$.
For $\Lambda\to 0$ one finds in the flat phase a Poisson's ratio 
$\sigma_f\approx -1/3$, as in the NPRG derivative 
expansion \cite{Kownacki09},
and in 
perfect agreement with both the SCSA result \cite{Doussal92} and
MC results for phantom membranes \cite{Falcioni97} and in good agreement with
MC results for self-avoiding membranes \cite{Bowick01}.

\begin{figure}[ht]
\begin{center}
\includegraphics[width=6.1cm,angle=-90]{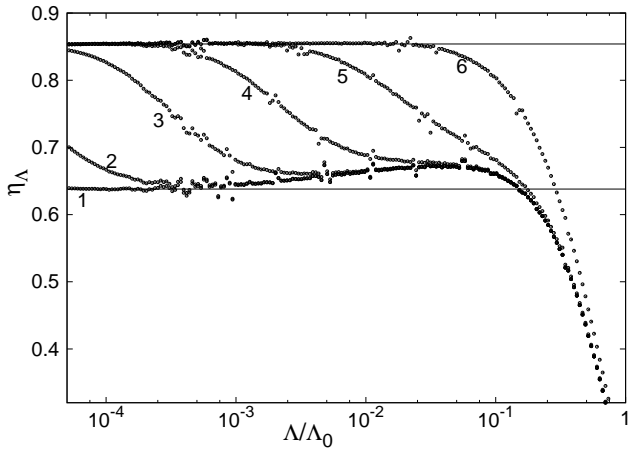}
\end{center}
\caption{Flow of the critical exponent $\eta$ for different initial values of 
$J_{\Lambda_0}^2$ (solid dots). For finite $J^2=\lim_{\Lambda \to 0} J_\Lambda^2$,
$\eta_\Lambda$ saturates at the flat phase fixed point value $\eta_f\approx 0.85$, this value
is indicated by the upper vertical line. 
At the crumpling transition (with $J_{\Lambda_0}=J_c$) $\eta_\Lambda$
saturates at the critical fixed point value $\eta_c\approx 0.638$ (see curve 1), this value is 
 indicated by the lower vertical line. 
The values for
$(J_{\Lambda_0}^2-J_c^2)/J_c^2$ are 
 $8.0 \times 10^{-7}$ (curve 2), $3.0 \times 10^{-5}$ (curve 3), $5.0\times  10^{-4}$ (curve 4), $9.1\times 10^{-3}$ (curve 5),
and $1.9\times 10^{-1}$ (curve 6).
} \label{fig:eta}
\end{figure}

\subsection{Behavior close to and at the crumpling transition}

Upon lowering the initial value of $J_{\Lambda_0}^2$, we can tune the
membrane towards the critical regime of the crumpling transition. 
Exactly at the crumpling transition the value of the anomalous
dimension is changed from the flat phase fixed point value.
The lowest curve in Fig.~\ref{fig:eta} shows the flow of $\eta_\Lambda$
at the critical value of $J_{\Lambda_0}^2$. One can clearly
observe a fixed point
value $\eta_c$ which is different from the one obtained for the flat
phase. We find a value 
\begin{equation}
  \eta_c\approx 0.63(8) \, ,
  \label{eq:etac}
\end{equation}
which is larger than the SCSA result $\eta_c= 0.535$ \cite{Doussal92},
and the value $\eta_c=0.47$ from a MC renormalization group  
analysis \cite{Espriu96}, but
rather close to the large $D$ result $\eta= 2/3$ and the
derivative expansion approach to the NPRG \cite{Kownacki09} 
$\eta\approx 0.627$, obtained with a sharp cutoff. 
In the derivative
expansion, a weak dependence of the critical properties on the
form of the regulator was reported. In our numerical approach,
we are for reasons of numerical stability restricted to analytical regulators
\cite{note1}.
We would expect however a much weaker dependence on the form of the
regulator in our approach 
since we keep the entire momentum dependence
of all two-point vertices and since the regulator, for a given $\Lambda$, only 
affects the momentum dependence of the two-point 
functions. 

\begin{figure}[ht]
\begin{center}
\includegraphics[width=6.1cm,angle=-90]{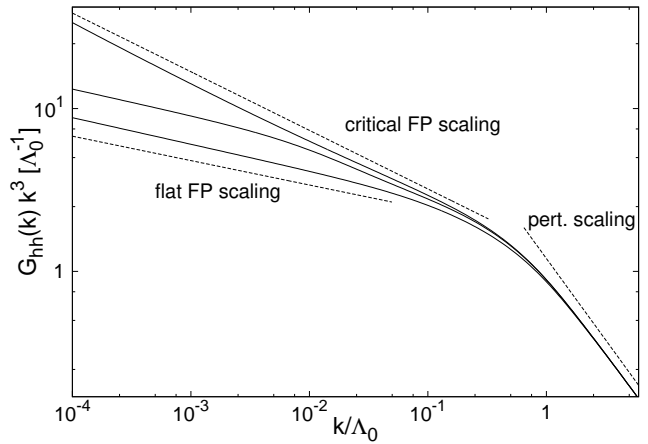}
\end{center}
\caption{The correlation function $G_{hh}(k)$ (solid lines) for different
values of $J_{\Lambda_0}^2-J_{c}^2$ (curves are multiplied with $k^3$ for
easier visibility of the different scaling regimes).
 At the critical point of the crumbling transition (upmost curve) a single
crossover from the perturbative scaling to a critical scaling with 
$\eta_c \approx 0.63(8)$ is observed. Slightly away from criticality (middle
curve),
$G_{hh}$ enters at small momenta the flat phase fixed point scaling
with $\eta_{f} \approx 0.85$. Even further away (lower curve), the
intermediate regime, where $G_{hh}$ obeys critical scaling, disappears
and a direct crossover from perturbative to flat phase fixed point scaling is
observed. Dashed lines indicate the different scaling behaviors.
} \label{fig:ghh}
\end{figure}

A small distance $J_{\Lambda_0}^2-J_c^2>0$ away from the critical 
value $J_c^2$,
$\eta_\Lambda$ initially approaches the critical fixed point value
$\eta_c$ but deviates from it in the IR limit to finally saturate
at the flat phase fixed point. This behavior can be seen for different
values of $J_{\Lambda_0}^2>J_c^2$ in Fig.~\ref{fig:eta}. 

The behavior of the Green's function $G_{hh}(q)$ at the critical
point is shown in Fig.~\ref{fig:ghh}. There is again a crossover
from perturbative scaling to anomalous scaling near $q_G$, similar
as in the flat phase. The anomalous scaling is now the critical
scaling with $\eta=\eta_c$. As expected from an analysis of the
flow of the critical exponent, there is a second crossover at
a smaller scale $q_c$ at which the critical scaling regime
terminates and scaling associated with the flat phase fixed
point critical exponent $\eta_f$ appears. A similar two-parameter
scaling is in fact also present in $\Phi^4$ models at
weak coupling \cite{Hasselmann07}.

The scale $q_c$ can be estimated from the
simple observation that the crossover occurs for 
$\tilde{\kappa}_{q_c} q_c^2\approx J^2 \tilde{\mu}_{q_c}$.
Since 
$\tilde{\kappa}_{q}\propto q^{-\eta_c}$ and
$\tilde{\mu}_q\propto q^{4-D-2 \eta_c}$, we find for $D=2$
the relation
\begin{equation}
  \label{eq:qc}
  q_c\propto J^{2/\eta_c} \, ,
\end{equation}
where $J$ is the fully renormalized magnitude of the order 
parameter which vanishes
at the critical point.
\begin{figure}[ht]
\begin{center}
\includegraphics[width=6.1cm,angle=-90]{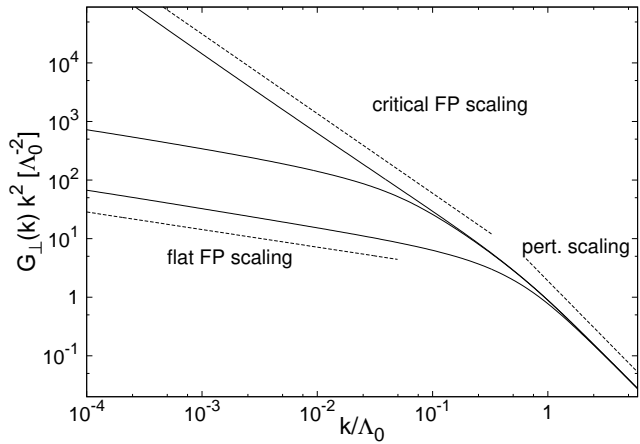}
\end{center}
\caption{The correlation function $G_{\perp}(k)$ (solid lines) for different
values of $J_{\Lambda_0}^2-J_{c}^2$ (curves are multiplied with $k^2$ for
easier visibility of the different scaling regimes).
 At the critical point of the crumbling transition (upmost curve) a single
crossover from the perturbative scaling to a critical scaling with 
$\eta_c \approx 0.63(8)$ is observed. At criticality, we have 
$G_{hh}(k)= G_\perp(k)=G_\parallel(k)$ since the order parameter
$J$ vanishes.
Slightly away from criticality (middle
curve),
$G_{hh}$ enters at small momenta the flat phase fixed point scaling
with $\eta^u=4-D-2 \eta_{f} \approx 0.30$. Even further away (lower curve), the
intermediate regime, where $G_{hh}$ obeys critical scaling, disappears
and a direct crossover from perturbative to flat phase fixed point scaling is
observed. Dashed lines indicate the different scaling behaviors.
} \label{fig:gperp}
\end{figure}

The in-plane fluctuations show a similar behavior, 
see Fig.~\ref{fig:gperp} (we only show results for $G_\perp(k)$,
but $G_\parallel(k)$ shows a very similar behavior). Directly at the
crumpling transition the order parameter $J$ vanishes and one therefore 
has $G_{hh}(q)=G_\perp(q)=G_\parallel(q)$ for all $q$. Similarly,
a finite distance away from the critical surface, but for momenta
in the regime $q_c\ll q \ll q_G$ the fluctuations are
still dominated by the vicinity of the critical fixed point and consequently
the in-plane modes scale anomalously with the same exponent as
$G_{hh}$, i.e., with $\alpha=\perp,\parallel$, 
\begin{equation}
G_{\alpha}(q) \simeq G_{hh} (q) \propto \frac{1}{q^{4-\eta_c}}
\, \, \, \,
\mbox{for $q_c\ll q \ll q_G$.}
\end{equation}
On the other hand, for very small $q$ one obtains
\begin{equation}
  G_{\alpha }\propto \frac{1}{q^{2+\eta^u}} \, \, \, \,
\mbox{for $q\ll q_c $,}
\end{equation}
as expected, since at small $\Lambda$ (or small $q$), the NPRG flow is away 
from the crumpling transition fixed point and towards the flat phase fixed point.
Note that the crossover at $q_c$ takes the in-plane modes thus directly
from a $q^{-4+\eta_c}$ scaling to a $q^{-2-\eta^u}$ scaling. Thus,
a regime where the in-plane modes scale with an anomalous dimension
$4-D-2\eta_c$ is not present. While we do observe a 
$\tilde{\mu}_q\sim \tilde{\lambda}_q \sim q^{\eta_c^u}$ scaling with
$\eta_c^u={4-D-2\eta_c}$ in $D=2$,
this exponent is not observable in the in-plane correlation functions
because $J_\Lambda^2\sim (\Lambda/\Lambda_0)^{D-2+\eta_c}$ on
approaching the crumpling transition such that the contribution to the
self-energies via  
$J_\Lambda^2 \tilde{\mu}_q q^2\sim J_\Lambda^2 \tilde{\lambda}_q q^2\sim q^{4-\eta_c}$
 (with $\Lambda\sim q$)
scale with the same exponent as $\tilde{\kappa}_q q^4$.
\begin{figure}[ht]
\includegraphics[width=6.cm,angle=-90]{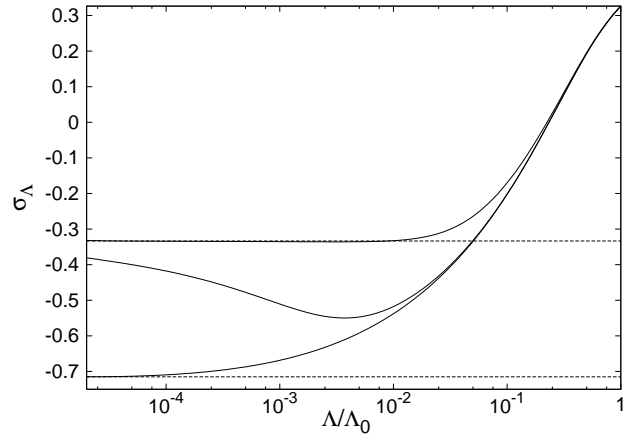}
\caption{Flow of the Poisson's ratio $\sigma_\Lambda$. In the
flat phase, we find $\sigma=-1/3$ whereas at the
crumpling transition we find a much larger magnitude, $\sigma\approx -0.71(5)$.
Both fixed point values for $\sigma_\Lambda$ are indicated by
dashed lines. 
The flow in the middle shows a flow of $\sigma_\Lambda$ for a membrane close to the critical
point, with the IR properties however dominated by the flat phase fixed point.
}
\label{fig:poisson}
\end{figure}

Rather interesting is the behavior of the Poisson's ratio, defined
in Eq.~(\ref{eq:poisson}), at the crumpling transition, see Fig.~\ref{fig:poisson}. It is more than
twice the value found for the flat phase,
\begin{equation}
  \sigma_c \approx -0.71(5) \, .
\end{equation}
The smallest possible value of $\sigma$ is $-1$ which is achieved when the 
ratio of the bulk modulus to the shear modulus vanishes,
i.e. $\lim_{k\to 0} (\tilde{\mu}_k+\tilde{\lambda}_k)/\tilde{\mu}_k \to 0$.
The large negative value of $\sigma_c$ thus implies that at criticality
the bulk modulus is very small compared to the dominant shear modulus. 
While several other approaches (e.g. the SCSA \cite{Doussal92} or the
derivative expansion approach to the NPRG \cite{Kownacki09}) would allow
to extract the value of the Poisson's ratio at criticality, we are
not aware of any earlier results. If the crumpling transition
is indeed of second order, its Poisson's ratio would be comparable
to those of the most strongly auxetic materials (materials with a negative
Poisson's ratio) presently known \cite{Lakes87}. 
The lack of orientational order at criticality, and the resulting 
absence of a $D=2$-dimensional plane along which the
membrane extends, complicate however the physical interpretation
of a large negative Poisson's ratio calculated for $D=2$.
The flow of $\sigma_\Lambda$
is shown in Fig.~\ref{fig:poisson} as a function of the IR cutoff $\Lambda$.
The lowest curve is the flow towards the fixed point of the crumpling transition.
Also shown is the flow of the Poisson's ratio for parameters close to the
critical ones but where the flow is ultimately to the value $\sigma_f=-1/3$ associated
with the flat phase.

\begin{figure}[ht]
\includegraphics[width=6.cm,angle=-90]{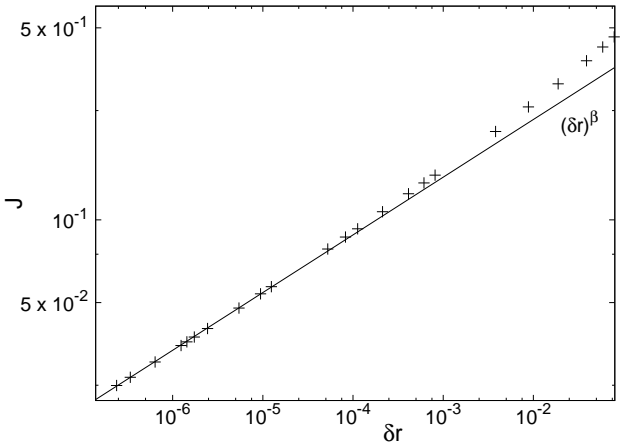}
\caption{Dependence of the fully renormalized order parameter
$J$ as a function of $\delta r=\tilde{r}_c-\tilde{r}_0$. The critical
exponent is found to be $\beta \approx 0.22$.}
\label{fig:beta}
\end{figure}

The magnitude of the order parameter $J$ scales near the crumpling transition
with an exponent $\beta$,
\begin{equation}
  \label{eq:betascal}
  J \simeq (\delta{r})^\beta \, ,
\end{equation}
where $\delta {r}=\tilde{r}_c-\tilde{r}_0>0$ measures the distance to the critical temperature,
$\delta r \propto (T_c-T)/T_c$. Mean field theory predicts $\beta=1/2$, see Eq.~(\ref{eq:Jmf}),
while our NPRG calculation yields a significantly different value for $\beta$,
\begin{equation}
  \beta\approx 0.22 \, ,
\end{equation}
which was obtained from a fit for small values of $J$, see 
Fig.~\ref{fig:beta}. Note that from Eq.~(\ref{eq:qc}) and Eq.~(\ref{eq:betascal})
one finds a simple relation between the crossover scale $q_c$ and the distance
to the crumpling transition critical point,
\begin{equation}
  q_c \propto (\delta r)^\frac{2\beta}{\eta_c} \, .
\end{equation}
This is just the usual hyperscaling relation $\beta=\nu(D-2+\eta_c)/2$ \cite{Aronovitz89}
for $D=2$
where $\nu$ is the thermal exponent characterizing the divergence of the correlation length, 
with $\xi^{-1}\propto  (\delta r)^\nu$.
Identifying $q_c\approx \xi^{-1}$, we can deduce
\begin{equation}
  \nu\approx 0.69 \, .
\end{equation}
This value is in reasonable agreement with numerical works which find
$\nu \approx 0.85$ \cite{Espriu96} and $\nu=0.71(5)$ \cite{Wheater96} (older
results on $\nu$ can be found in Ref.~\cite{Espriu96}), but is 
somewhat larger than the value found in the derivative expansion of the NPRG
$\nu\approx  0.52$ \cite{Kownacki09}.
When approaching the crumpling transition from the ordered side,
 the in- and out-of-plane 
correlation functions do not decay algebraically for 
distances $ r \gg \xi$
with a characteristic scale $\xi$ but merely decay with a 
different power-law. This is why we refer to this scale as a crossover scale rather than a correlation
length.

While our analysis shows that the crumpling transition is of second
order, we based our analyis on the assumption that initially, at
the UV cutoff $\Lambda_0$, all coupling functions are momentum
independent. We ran a few tests to verify that our results are stable 
also in presence of a weak initial momentum dependence, but
we
cannot rule out that an unusual initial momentum dependence
of the coupling constants could lead to an instability, e.g. at
a finite momentum, in the renormalized model which would lead
to a first order transition. It is also possible that
including terms of third order in the stress tensor could 
modify the order of the transition.

\section{Conclusions}
\label{sec:conclusion}
To conclude, we have presented a thorough analysis of crystalline
phantom membranes using a NPRG scheme which includes the full momentum
dependence of the elastic coupling functions. It is a natural
extension of the NPRG scheme based on a derivative expansion
\cite{Kownacki09} but yields significantly more information
on the nature of the fluctuations. Since conflicting results
on the order of the crumpling transition exist, it is important
to include as many correlations as possible in the ansatz for
the effective average action. A-priori, it is difficult to
decide which terms will be relevant near the transition since
the anomalous dimension of the crumpling transition is very large
and the crumpling transition could even be of weakly first
order. Our nonlocal ansatz goes well beyond previous renormalization
group treatments of
crystalline membranes which relied on a finite number of
coupling parameters and should thus yield more reliable results.

In our approach we find a continuous crumpling transition for physical
membranes, i.e. $D=2$ dimensional membranes embedded in $d=3$ dimensional
space, and
we compute the associated critical exponents. We find an anomalous
dimension $\eta_c \approx 0.63(8)$ and a thermal exponent $\nu\approx 0.69$.
An analysis of the scaling of the renormalized order parameter near the crumpling
transition yields the critical exponent $\beta\approx 0.22$.
Inside the flat phase we find an anomalous dimension $\eta_f\approx 0.85$
which characterizes the asymptotic small momentum behavior of the out-of-plane
fluctuations and an additional anomalous dimension $\eta^u=4-D-2\eta_f\approx 0.30$ which
characterizes the asymptotic behavior of the in-plane fluctuations. 
We further analysed in detail the momentum dependence of the thermal
fluctuations of the membrane at finite momenta. In both the flat phase
and at the crumpling transition there is a crossover scale
$q_G$ which separates the anomalous scaling regime at small momenta from the perturbative 
regime. Near the crumpling transition, there is an additional crossover momentum scale
$q_c \propto [(T-T_c)/T_c]^\nu$ 
which separates an intermediate scaling regime, whose properties are determined
by the crumpling transition fixed point, from the asymptotic small scaling regime
where the flow is dominated by the flat phase fixed point.

We further calculated the Poisson's ratio both at the crumpling transition
and inside the flat phase. Inside the flat phase we recover the value $\sigma_f\approx -1/3$
whereas we find a Poisson's ratio of much large magnitude at the crumpling 
transition, $\sigma_c\approx -0.71(5)$. 

\acknowledgments 
We thank Dominique Mouhanna, Christoph Husemann, and Walter Metzner 
for discussions and comments.
\appendix*
\section{Transverse and longitudinal projection of the diagram $S_{ab}^{uu}(k)$}
Here we present the expression for the transverse and longitudinal projection
of the diagram $S_{ab}^{uu}(k)$ given in Eq.~(\ref{eq:Sabuu}), which enters
the NPRG flow equations of $\Sigma_{ab}(k)$. While straightforward to evaluate,
they have a complicated structure.
The projections of $S_{ab}^{uu}(k)$ can be written as:
\begin{subequations}
  \begin{align}  
    S_{\perp}^{uu} &=
    J_\Lambda^2  \int_{{\bm q}}
    \Big(  D_{\perp}^{\perp\perp}  
    \dot{G}_{\perp} (q)  {G}_{\perp} ({q^\prime})  + 
    D_{\perp}^{\parallel\parallel}
    \dot{G}_{\parallel} (q)  {G}_{\parallel}  ({q^\prime}) 
      \nonumber \\ & \qquad
    + D_{\perp}^{\perp \parallel}  
    \dot{G}_{\perp} (q)  {G}_{\parallel} ({q^\prime}) 
    + D_{\perp}^{\parallel\perp}  
    \dot{G}_{\parallel} (q)  {G}_{\perp} ({q^\prime}) 
    \Big) \, ,
    \\
    S_{\parallel}^{uu}  &=
    J_\Lambda^2
    \int_{{\bm q}}
    \Big(D_{\parallel}^{\perp\perp}  
      \dot{G}_{\perp} (q)   {G}_{\perp} ({q^\prime}) 
      + D_{\parallel}^{\parallel\parallel}  
      \dot{G}_{\parallel} ({\bm q})  {G}_{\parallel} ({q^\prime}) 
      \nonumber \\ & \qquad
      +D_{\parallel}^{\perp\parallel}  
      \dot{G}_{\perp} (q)  {G}_{\parallel} ({q^\prime}) 
      +D_{\parallel}^{\parallel \perp}   
      \dot{G}_{\parallel} (q)  {G}_{\perp}({q^\prime})  \Big) \, ,
    \end{align}
\end{subequations}
where we defined $\bd{q}^\prime=\bd{k}+\bd{q}$. If we further introduce

\begin{equation}
    X = {\bm q} \cdot {\bm k} \, , \, \, \,
    Y = {\bm q} \cdot \bd{q}^\prime \, , \, \, \,
    Z = {\bm k} \cdot \bd{q}^\prime \, ,
\end{equation}%
the functions $D_{\alpha}^{\beta\gamma}$ with 
$\alpha,\beta,\gamma=\parallel,\perp$ can be written as
\begin{widetext}
  \begin{subequations}
    \begin{align}
      D_{\perp}^{\perp \perp} &= 
            \tilde{\mu}_{{q^\prime}}^2 \left(k^2 Y^2 + q^2 Z^2 + 2 X\, Y Z 
        - 4 \frac{Y^2 Z^2}{{q^\prime}^2} \right) \left(D - 2 + 
        \frac{X^2}{k^2 q^2} \right) 
      + 2 \tilde{\mu}_{{q^\prime}}  \tilde{\mu}_{q} X Z 
      \left(1 - \frac{2Y}{q^2}\right) 
      \left(q^2 - \frac{X^2}{k^2}\right)
      \left(1 - 2 \frac{Y}{{q^\prime}^2}\right)
      \nonumber
      \\
      &  \quad
      + 2 \tilde{\mu}_{{q^\prime}} \tilde{\mu}_{k}  \left( X + Z \right) 
      Y
      \left(- \frac{X}{q^2} + 2 \frac{YZ}{q^2 {q^\prime}^2 }  - 
        \frac{2 Z}{{q^\prime}^2} \right) 
      \left(q^2 - \frac{X^2}{k^2}\right)
      + \tilde{\mu}_{k}^2 \left(D - 2 + \frac{Y^2}{q^2 {q^\prime}^2}\right) 
      \left( X + Z\right)^2 \left({q^\prime}^2 - \frac{Z^2}{k^2}\right)  
      \nonumber
      \\
      &  \quad
      + \tilde{\mu}_{q}^2 \left[ Y^2 \left(k^2 - \frac{4X^2}{q^2}\right) + 
        2 X Y Z + X^2 {q^\prime}^2   \right]
      \left( D - 2 + \frac{Z^2}{k^2 {q^\prime}^2 } \right) 
      \nonumber
      \\
      &  \quad
      + 
      2 \tilde{\mu}_{k} \tilde{\mu}_{q}  Y\left(X + Z\right) 
      \left( -2 \frac{X }{q^2} 
        +2 \frac{X Y}{q^2 {q^\prime}^2} 
        - \frac{Z}{{q^\prime}^2} 
      \right) 
      \left(q^2 - \frac{X^2}{k^2}\right)  \, , 
      \\
      D_{\perp}^{\parallel\parallel} &=
      4\tilde{\mu}_{{q^\prime}}^2 \frac{Y^2 Z^2}{{q^\prime}^2} \left(1 - 
        \frac{X^2}{q^2 k^2} \right)
      + \tilde{\lambda}_{q^\prime}^2 X^2 {q^\prime}^2 
      \left(1 - \frac{X^2}{k^2 q^2}\right) 
      + 4 \tilde{\mu}_{{q^\prime}} \tilde{\lambda}_{{q^\prime}}  X Y Z 
      \left(1 - \frac{X^2}{k^2 q^2}\right) 
      + 8 \tilde{\mu}_{q} \tilde{\mu}_{q^\prime}  \frac{Z Y^2 X}{q^2{q^\prime}^2} 
      \left(q^2 - \frac{X^2}{k^2}\right) 
      \nonumber
      \\
      & \quad
      + 2 \tilde{\lambda}_{q} \tilde{\lambda}_{q^\prime}  X Z 
      \left(q^2 - \frac{X^2}{k^2}\right)  
      +4  \tilde{\mu}_{q} \tilde{\lambda}_{q}  X Y Z 
      \left(1 - \frac{Z^2}{k^2 {q^\prime}^2}\right) 
      +4 \tilde{\mu}_{q} \tilde{\lambda}_{q^\prime}  X^2 Y 
      \left(1 - \frac{X^2}{q^2 k^2} \right)  
      +4 \tilde{\mu}_{{q^\prime}} \tilde{\lambda}_{q}  
      \frac{Y Z^2}{{q^\prime}^2} \left(q^2 - \frac{X^2}{k^2}\right) 
      \nonumber
      \\
      & \quad
      +4 \tilde{\mu}_{k} \tilde{\mu}_{q^\prime}  \frac{Z Y^2}{{q^\prime}^2} \left(X + Z\right) 
      \left(1 - \frac{X^2}{q^2 k^2}\right)
      + 2 \tilde{\mu}_{k} \tilde{\lambda}_{q^\prime} \left(X + Z\right) X Y 
      \left(1 - \frac{X^2}{q^2 k^2}\right) + 
      4 \tilde{\mu}_{q}^2  \frac{X^2 Y^2}{q^2} 
      \left(1 - \frac{Z^2}{k^2 {q^\prime}^2} \right)
      \nonumber
      \\
      & \quad
      + \tilde{\lambda}_{q}^2 Z^2 q^2 \left(1 - 
        \frac{Z^2}{k^2 {q^\prime}^2}\right)  
      + 4 \tilde{\mu}_{k} \tilde{\mu}_{q} \left(X + Z\right)  
      \frac{X Y^2}{{q^\prime}^2} \left(1 - \frac{X^2}{q^2 k^2}\right) 
      + 
      2 \tilde{\mu}_{k} \tilde{\lambda}_{q} \left(X + Z\right) \frac{Z Y}{{q^\prime}^2} 
      \left(q^2 - \frac{X^2}{k^2}\right)
      \nonumber
      \\
      & \quad
      +  \tilde{\mu}_{k}^2 \left(X + Z\right)^2 \frac{Y^2}{{q^\prime}^2} 
      \left(1 - \frac{X^2}{q^2 k^2}\right) \, ,
      \\
      D_{\perp}^{\perp \parallel} &= 
      \frac{1}{{q^\prime}^2}\left(2 \tilde{\mu}_{{q^\prime}} Y Z +  
        \tilde{\lambda}_{{q^\prime}} X {q^\prime}^2 \right)^2 
      \left(D - 2 + \frac{X^2}{k^2 q^2}\right)
      + 4 \tilde{\mu}_{q} \tilde{\mu}_{{q^\prime}}  \frac{X Y Z}{{q^\prime}^2} 
      \left(1 - 2 \frac{Y}{q^2}\right) \left(q^2 - \frac{X^2}{k^2}\right)
      + 2 \tilde{\mu}_{q} \tilde{\lambda}_{{q^\prime}} X^2 \left(1 - 2 \frac{Y}{q^2}\right)
      \nonumber
      \\
      & \quad 
      \times \left(q^2 - \frac{X^2}{k^2}\right)
      + 4 \tilde{\mu}_{k} \tilde{\mu}_{{q^\prime}} \left(X + Z\right)  
      \frac{Y Z}{{q^\prime}^2} \left(1 - \frac{Y}{q^2}\right)
      \left(q^2 - \frac{X^2}{k^2}\right)
      + \tilde{\mu}_{q}^2 \biggl[
        Y^2 \left(k^2 - \frac{X^2}{q^2}\right) + 
        2 X Y \left(Z - \frac{Y X}{q^2}\right) 
      \nonumber
      \\
      & \quad
        + 
        X^2 \left({q^\prime}^2 - \frac{Y^2}{q^2}\right)\biggl] 
      \left(1 - \frac{Z^2}{k^2 {q^\prime}^2}\right)
      + 2  \tilde{\mu}_{k} \tilde{\mu}_{q} \left( X + Z\right) 
      \biggl[ X \left(1 - \frac{2Y^2}{q^2 {q^\prime}^2}\right) + 
        \frac{YZ}{{q^\prime}^2}   \biggl] 
      \left(q^2 - \frac{X^2}{k^2}\right)
      \nonumber
      \\
      & \quad
      + \tilde{\mu}_{k}^2 \left(X + Z\right)^2 \left(q^2 - \frac{X^2}{k^2}\right) 
      \left(1 - \frac{Y^2}{q^2 {q^\prime}^2}\right)   
      + 2 \tilde{\mu}_{k} \tilde{\lambda}_{{q^\prime}} \left(X + Z\right) X 
      \left(1 - \frac{Y}{q^2}\right) 
      \left(q^2 - \frac{X^2}{k^2}\right) \, ,
    \end{align}
    \begin{align}
      D_{\perp}^{\parallel\perp} &=
      \tilde{\mu}_{{q^\prime}}^2 \left(k^2 Y^2 + q^2 Z^2 + 2 X Y Z - 
        4 \frac{Y^2 Z^2}{{q^\prime}^2} \right) \left(1 - 
        \frac{X^2}{k^2 q^2}\right) 
      + 4 \tilde{\mu}_{q} \tilde{\mu}_{{q^\prime}}  X Y Z 
      \left(1 - 2 \frac{Y}{{q^\prime}^2}\right) 
      \left(1 - \frac{X^2}{k^2 q^2}\right)
      \nonumber
      \\
      & \quad 
      + 2 \tilde{\mu}_{{q^\prime}} \tilde{\lambda}_{q}  Z^2 
      \left(1 - 2 \frac{Y}{{q^\prime}^2}\right) 
      \left(q^2 - \frac{X^2}{k^2}\right) 
      + 2 
      \tilde{\mu}_{k} \tilde{\mu}_{{q^\prime}} \left(X + Z\right) \left( 
        \frac{Y X}{q^2} + Z - 2 \frac{Y^2 Z}{q^2 {q^\prime}^2} \right) 
      \left(q^2 - \frac{X^2}{k^2}\right)
      \nonumber
      \\
      & \quad 
      + 4 \tilde{\mu}_{q}^2  X^2 \frac{Y^2}{q^2}  
      \left( D - 2 + \frac{Z^2}{k^2 {q^\prime}^2}\right)
      + \tilde{\lambda}_{q}^2 Z^2 q^2 \left(D - 2 + 
        \frac{Z^2}{{q^\prime}^2  k^2}\right)
      +4 \tilde{\lambda}_{q} \tilde{\mu}_{q}  X Y Z \left(D - 2 + 
        \frac{Z^2}{k^2 {q^\prime}^2}\right)
      \nonumber
      \\
      & \quad 
      + 4 \tilde{\mu}_{q}  \tilde{\mu}_{k} \left(X + Z\right)  X Y 
      \left(1 - \frac{Y}{{q^\prime}^2}\right) 
      \left(1 - \frac{X^2}{k^2 q^2}\right)
      + 2 \tilde{\lambda}_{q} \tilde{\mu}_{k} \left(X + Z\right) Z 
      \left(1 - \frac{Y}{{q^\prime}^2}\right) 
      \left(q^2 - \frac{X^2}{k^2}\right)   
      \nonumber
      \\
      & \quad 
      + \tilde{\mu}_{k}^2 \left(X + Z\right)^2 \left(q^2 - \frac{X^2}{k^2}\right) 
      \left(1 - \frac{Y^2}{q^2 {q^\prime}^2}\right) \, .
    \end{align}
  \end{subequations}
For the terms of the longitudinal projection one finds
  \begin{subequations}
    \begin{align}
      D_{\parallel}^{\perp\perp} &=
      \tilde{\mu}_{{q^\prime}}^2 
      \left[ Y^2 
        \left(
          k^2 - \frac{Z^2}{{q^\prime}^2}
        \right) + 2 Y Z 
        \left(
          X - \frac{Y Z}{{q^\prime}^2}
        \right) + 
        Z^2 \left(
          q^2 - \frac{Y^2}{{q^\prime}^2}
        \right)
      \right] 
      \left(
        1 - \frac{X^2}{k^2 q^2} 
      \right)  
      +2  \tilde{\mu}_{q} \tilde{\mu}_{{q^\prime}} 
      \biggl[ Y \left(
        k^2 - \frac{Z^2}{{q^\prime}^2}
      \right) 
      \nonumber
      \\
      & \quad
      + 
      Z \left(
        X - \frac{Y Z}{{q^\prime}^2} 
      \right) \biggr] 
      \left[ 
        \frac{X}{k^2} \left(
          Z - \frac{X Y}{q^2}
        \right) + 
        Y \left(
          1 - \frac{X^2}{k^2 q^2}
        \right) 
      \right]
      + 
      2 \tilde{\mu}_{q^\prime} Y 
      \left(2 \tilde{\mu}_{k} \frac{X Z}{k^2} + 
        \tilde{\lambda}_{k}Y \right) 
      \biggl(k^2  - \frac{X^2}{q^2} - 
      2 \frac{ Z^2}{{q^\prime}^2}         
      +2 
      \frac{X Y Z  }{q^2 {q^\prime}^2} \biggr)
      \nonumber
      \\
      & \quad 
      + 
      \tilde{\mu}_{q}^2 \left[ X^2 
        \left({q^\prime}^2 - \frac{Y^2}{q^2}\right)  + 
        Y^2 \left(k^2 - \frac{X^2}{q^2}\right) + 
        2 X Y  \left(Z - \frac{X Y}{q^2}\right)  \right]
      \left(1 - \frac{Z^2}{{q^\prime}^2  k^2} \right) 
      +2\tilde{\mu}_{q} Y  
      \left( 2\tilde{\mu}_{k}  \frac{X Z}{k^2} + \tilde{\lambda}_{k} 
        Y 
      \right)
      \nonumber
      \\
      & \quad 
      \times \left(
        k^2- \frac{2X^2}{q^2} + \frac{2XYZ}{q^2 {q^\prime}^2}
        - \frac{Z^2}{{q^\prime}^2}
      \right)
      + \left(
        D - 2 + \frac{Y^2}{q^2 {q^\prime}^2} 
      \right) 
      \left(
        2 \frac{X Z}{k^2} \tilde{\mu}_{k} + 
        Y \tilde{\lambda}_{k} 
      \right)^2 k^2 
      \, , 
      \\
      D_\parallel^{\parallel\parallel}&=  
            4 \tilde{\mu}_{q^\prime}( \tilde{\mu}_{q^\prime}+2\tilde{\mu}_q) 
      \frac{X^2 Y^2 Z^2 }{k^2 q^2 {q^\prime}^2} + 
      4\tilde{\mu}_{{q^\prime}}  \tilde{\lambda}_{{q^\prime}} 
      \frac{X^3 Y Z}{k^2 q^2} 
      + \tilde{\lambda}_{{q^\prime}}^2  
      \frac{X^4{q^\prime}^2}{k^2 q^2} 
      + 
      2 \tilde{\lambda}_{{q^\prime}} \tilde{\lambda}_{q} 
      \frac{X^2 Z^2}{k^2}   
      + 4 \tilde{\mu}_{{q^\prime}}  \tilde{\lambda}_{q}    
      \frac{X Y Z^3}{k^2 {q^\prime}^2} +
      4 \tilde{\lambda}_{{q^\prime}} \tilde{\mu}_{q} 
      \frac{ X^3 Y Z}{q^2 k^2}
      \nonumber
      \\
      & \quad
      + 
      \frac{4XY}{q^2 k^2 {q^\prime}^2} \left( \tilde{\mu}_{k}  X Z + 
        \tilde{\lambda}_{k} Y k^2\right) \left( 2 Y Z 
        \tilde{\mu}_{{q^\prime}} + X  {q^\prime}^2 
        \tilde{\lambda}_{{q^\prime}}\right)
      +  \frac{Z^2}{k^2 q^2 {q^\prime}^2} 
      \left(
        2 \tilde{\mu}_{q} X Y
        + \tilde{\lambda}_{q} q^2 Z 
      \right)^2 
      \nonumber
      \\
      & \quad 
      + 
      \frac{2YZ}{k^2  {q^\prime}^2} \left( 2\tilde{\mu}_{k} X Z + 
        \tilde{\lambda}_{k} Y k^2\right)  
      \left(2 \frac{X Y}{q^2} \tilde{\mu}_{q} + Z  
        \tilde{\lambda}_{q}\right)  
      + \left( 2\tilde{\mu}_{k} X Z + \tilde{\lambda}_{k} Y k^2\right)^2
      \frac{Y^2}{k^2 q^2 {q^\prime}^2} \, , 
    \end{align}
    \begin{align}
      D_{\parallel}^{\perp\parallel} &=
      2 \tilde{\lambda}_{{q^\prime}} \frac{X}{k^2} 
      \left(Z - \frac{X Y}{q^2}\right) \left( 2 \tilde{\mu}_{k}  X Z + 
        \tilde{\lambda}_{k} Y k^2\right) +
      \left(1 - \frac{X^2}{q^2 k^2}\right) 
      {q^\prime}^{-2}\left(  
        2\tilde{\mu}_{{q^\prime}}  Y Z 
        + \tilde{\lambda}_{{q^\prime}} {q^\prime}^2 X\right)^2 
      \nonumber
      \\
      & \quad 
      +4 \tilde{\mu}_{{q^\prime}} \tilde{\mu}_{q} 
      Y Z^2 \left[ X \left(Z - \frac{X Y}{q^2}\right) + 
        \frac{Y}{{q^\prime}^2 k^2} \left(k^2 - \frac{X^2}{q^2}\right)\right]
      +   2 \tilde{\lambda}_{{q^\prime}} \tilde{\mu}_{q} \frac{X}{k^2}  
      \left(X Z^2 - 2 \frac{X^2 Y Z}{q^2} + Y Z k^2\right) 
      \nonumber
      \\
      & \quad 
      + 
      4 \tilde{\mu}_{{q^\prime}} \frac{YZ}{{q^\prime}^2 k^2} 
      \left(Z -\frac{ X Y}{q^2}\right) 
      \left( 2 \tilde{\mu}_{k}  X Z + \tilde{\lambda}_{k} Y k^2\right) 
      +\frac{\tilde{\mu}_{q}^2 Z^2}{k^2 {q^\prime}^2} 
      \left[ X^2 \left({q^\prime}^2 - 
          \frac{Y^2}{q^2}\right) + Y^2 \left(k^2 - \frac{X^2}{q^2}\right) + 
        2 X Y  \left(Z - \frac{X Y}{q^2}\right) \right] 
      \nonumber
      \\
      & \quad 
      +2 \tilde{\mu}_{q} \frac{Z}{k^2} \left(2\tilde{\mu}_{k} X Z + 
        \tilde{\lambda}_{k} Y k^2\right) 
      \left[ X \left(1 - \frac{2 Y^2 }{q^2 {q^\prime}^2} \right) + 
        \frac{YZ }{{q^\prime}^2} \right]
      + 
      \left(2 X Z \tilde{\mu}_{k} + \tilde{\lambda}_{k} Y k^2\right)^2
      \left(1 - \frac{Y^2}{q^2 {q^\prime}^2}\right) k^{-2} \, , 
      \\
      D_\parallel^{\parallel\perp} &= 
      \frac{\tilde{\mu}_{{q^\prime}}^2 X^2}{k^2 q^2} 
      \left[Y^2 \left(k^2 - \frac{4Z^2}{{q^\prime}^2}
        \right)
        + 2 X Y Z +Z^2 q^2   \right] +
      2 \tilde{\mu}_{{q^\prime}}\frac{X}{k^2}
      \left(
        2 \tilde{\mu}_{q}\frac{XY}{q^2}
        +\tilde{\lambda_q} Z\right)
      \left[ Y \left(k^2 - \frac{2Z^2}{{q^\prime}^2}\right) + 
        Z X \right]  
      \nonumber
      \\
      & \quad 
      +      
      2  \frac{\tilde{\mu}_{q^\prime}X}{k^2 q^2} 
      \left( 2 \tilde{\mu}_{k} X Z
        + \tilde{\lambda}_{k} Y k^2\right) \left(XY + Z q^2 - 
        2 \frac{Z Y^2}{{q^\prime}^2}\right) 
      + q^{-2}
      \left( 
        2\tilde{\mu}_{q}  XY + 
        \tilde{\lambda}_{q} Z q^2 
      \right)^2
      \left(1 - \frac{ Z^2}{k^2 {q^\prime}^2}\right) 
      \nonumber
      \\
      & \quad 
      +
      2
      \left(  2\tilde{\mu}_{q} \frac{XY}{q^2}+\tilde{\lambda}_qZ
      \right)\left(X - 
        \frac{Y Z}{{q^\prime}^2}\right)   
      \left(2\tilde{\mu}_{k} \frac{X Z}{k^2}  + \tilde{\lambda}_{k}Y\right) 
       + \left( 2 \tilde{\mu}_{k} \frac{X Z}{k^2}
         + \tilde{\lambda}_{k}  Y \right)^2 
       \left(1 - 
         \frac{Y^2}{q^2 {q^\prime}^2} \right)k^2 \, .
    \end{align}
   \end{subequations}
\end{widetext}


\begin{thebibliography}{99}

\bibitem{membranebook} D.~Nelson, T.~Piran, and S.~Weinberg (Eds.), {\em Statistical Mechanics of Membranes
and Surfaces}, 2nd edition, World Scientific, Singapore (2004).
%
\bibitem{Nelson87} D. R. Nelson and L. Peliti, J. Phys. (Paris) {\bf 48}, 1085 (1987).
%
\bibitem{Castro09} See A. H. Castro Neto, F.~Guinea, N.~M.~R.~Peres, 
K.~S.~Novoselov, and A.~K.~Geim, Rev.~Mod.~Phys {\bf 81}, 109 (2009)
for a review.
%
\bibitem{Meyer07} J. C. Meyer, A. K. Geim, M. I. Katsnelson, K. S. Novoselov, T. J. Booth, and S. Roth,
Nature {\bf 446}, 60 (2007).
%
\bibitem{Paczuski88} M. Paczuski, M. Kardar, and D. R. Nelson, {Phys. Rev. Lett.} {\bf 60}, 2638 (1988).
%
\bibitem{Aronovitz89} J. Aronovitz, L. Golubovi\'c, and T. C. Lubensky, {J. Phys. (Paris)} {\bf 50}, 609 (1989).
%
\bibitem{Paczuski89} M. Paczuski and M. Kardar, {Phys. Rev. A} {\bf 39}, 6086 (1989).
%
\bibitem{Doussal92} P. Le Doussal and L. Radzihovsky, Phys. Rev. Lett. {\bf 69},
1209 (1992).
%
\bibitem{Gazit09} D. Gazit, Phys. Rev. E {\bf 80}, 041117 (2009).
%
\bibitem{Zakharchenko10} K.~V.~Zakharchenko, R.~Rold\'{a}n, A.~Fasolino, and M.~I.~Katsnelson,
 Phys. Rev. B {\bf 82}, 125435 (2010).
%
\bibitem{David88} F. David and E. Guitter, Europhys. Lett. {\bf 5}, 709 (1988).
%
\bibitem{Espriu96} D.~Espriu and A.~Travesset, Nucl.Phys.B {\bf 468}, 514-540 (1996).
%
\bibitem{Bowick01rev} M.~J.~Bowick and A.~Travesset, 
Phys.~Rep. {\bf 344}, 255 (2001).
%
\bibitem{Kownacki02}
J.~Ph.~Kownacki and H.~T.~Diep, Phys. Rev. E {\bf 66}, 066105 (2002).
%
\bibitem{Koibuchi04}
H.~Koibuchi, N.~Kusano, A.~Nidaira, K.~Suzuki, and M.~Yamada,
Phys. Rev. E {\bf 69}, 066139 (2004).
%
\bibitem{Koibuchi08}
H.~Koibuchi, Phys. Rev. E {\bf 77}, 021104 (2008).
%
\bibitem{Nishiyama10} 
Y.~Nishiyama, Phys. Rev. E {\bf 82}, 012102 (2010).
%
\bibitem{Kownacki09} J. P. Kownacki and D. Mouhanna, {Phys. Rev. E} {\bf 79}, 040101(R) (2009).
%
\bibitem{Essafi10} 
K.~Essafi, J.-P.~Kownacki, and J.~Mouhanna, preprint, arXiv:1011.6173 (2010).
%
\bibitem{Braghin10} F.L. Braghin and N. Hasselmann, {Phys. Rev. B} {\bf 82}, 035407 (2010).
%
\bibitem{Fasolino07} A. Fasolino, J. H. Los, and M. I. Katsnelson, Nature Mater. {\bf 6}, 858 (2007).
%
\bibitem{Los09} J. H. Los, M. I. Katsnelson, O. V. Yazyev, K. V. Zakharchenko, and A. Fasolino,
{Phys. Rev. B} {\bf 80}, 121405(R) (2009).
%
\bibitem{phi4} S. Ledowski, N. Hasselmann, and P. Kopietz, {Phys. Rev.} A {\bf 69}, 061601(R) (2004); N. Hasselmann, S. Ledowski, and P. Kopietz, 
{\em ibid.} {\bf 70}, 063621 (2004); J.-P. Blaizot, R. M\'{e}ndez-Galain, 
and N. Wschebor, 
Phys.~Rev.~E {\bf 74}, 051116 (2006);
A. Sinner, N. Hasselmann, and P. Kopietz, 
{J. Phys.: Cond. Mat.} {\bf 20}, 075208 (2008); 
F.~Benitez, J.-P.~Blaizot, H.~Chat\'{e}, B.~Delamotte, R.~M\'{e}ndez-Galain, and N.~Wschebor,
Phys. Rev. E {\bf 80}, 030103(R) (2009).
%
\bibitem{Sinner10} A. Sinner, N. Hasselmann, and P. Kopietz, Phys.~Rev.~A {\bf 82}, 063632 (2010).
%
\bibitem{Sinner09} A. Sinner, N. Hasselmann, and P. Kopietz, Phys. Rev. Lett. {\bf 102}, 120601 (2009);
N. Dupuis, Phys. Rev. Lett. {\bf 102}, 190401 (2009); N. Dupuis, Phys. Rev. A {\bf 80}, 043627 (2009).
%
\bibitem{Wetterich93} C. Wetterich, Phys. Lett. B {\bf 301}, 90 (1993); T. R. Morris,
Int. J. Mod. Phys. A {\bf 9}, 2411 (1994).
%
\bibitem{Berges02}
J.~Berges, N.~Tetradis, and C.~Wetterich, Phys.~Rep. {\bf 363} 223 (2002).
%
\bibitem{Schuetz06} F. Sch\"utz and P. Kopietz, J. Phys. A {\bf 39}, 8205 (2006).
%
\bibitem{Bowick96} M.J. Bowick, S.M. Catterall, M. Falcioni, G. Thorleifsson, 
and K.N. Anagnostopoulos, {J. Phys. (France)} {\bf 6}, 1321 (1996).
%
\bibitem{Zhang93} Z.~Zhang, H.~T.~Davis, and D.~M.~Kroll,
 Phys. Rev. E {\bf 48}, 651(R) (1993).
%
\bibitem{Falcioni97} M. Falcioni, M. J. Bowick, E. Guitter, and G. Thorleifsson,
Europhys.~Lett. {\bf 38}, 67 (1997). 
%
\bibitem{Bowick01} M. Bowick, A. Cacciuto, G. Thorleifsson, A. Travesset,
Phys.~Rev.~Lett. {\bf 87}, 148103 (2001).
%
\bibitem{note1} The value given in Eq.~(\ref{eq:etac}) is
actually slightly smaller than the value reported in \cite{Braghin10},
where the numerical calculation did not extend to sufficiently small
momenta. As can be seen in Fig.~\ref{fig:eta}, $\eta_\Lambda$ 
converges rather slowly at the crumpling transition to its fixed point 
value.
%
\bibitem{Hasselmann07}
N.~Hasselmann, A.~Sinner, and P.~Kopietz, Phys. Rev. E {\bf 76}, 040101(R) (2007).
%
\bibitem{Lakes87}  R. S. Lakes, Science {\bf 238}, 551 (1987).
%
\bibitem{Wheater96} J. F. Wheater, Nucl. Phys. B {\bf 458}, 671 (1996).

\end{thebibliography}
\end{document}